\documentclass[12pt,a4paper]{article}

\usepackage[english]{babel}
\usepackage{amsmath}
\usepackage[letterpaper,top=2cm,bottom=2cm,left=3cm,right=3cm,marginparwidth=1.75cm]{geometry}
\usepackage{titlesec}
\usepackage{multirow}
\usepackage{tabularx} 
\usepackage{booktabs}
\titleclass{\subsubsubsection}{straight}[\subsection]

\newcounter{subsubsubsection}[subsubsection]
\renewcommand\thesubsubsubsection{\thesubsubsection.\arabic{subsubsubsection}}
\titleformat{\subsubsubsection}
  {\normalfont\normalsize\bfseries}{\thesubsubsubsection}{1em}{}
\titlespacing*{\subsubsubsection}
{0pt}{3.25ex plus 1ex minus .2ex}{1.5ex plus .2ex}
\usepackage[style=apa, backend=biber,doi=false, url=false, maxcitenames=6]{biblatex}
\AtEveryCitekey{%
  \ifciteseen
    {\defcounter{maxnames}{2}}
    {}%
}
\usepackage{graphicx}
\usepackage{times}
\usepackage{booktabs}
\usepackage{algorithm}
\usepackage{algpseudocode}
\usepackage{amsmath}
\usepackage{amssymb}
\usepackage{setspace}
\usepackage{siunitx}
\usepackage[utf8]{inputenc}
\usepackage{xcolor}
\usepackage[colorlinks=true, allcolors=black]{hyperref}
\def\mathbi#1{\textbf{\em #1}}
\begin{document}

\begin{titlepage}
    \centering
    \vspace{1cm}
    {\normalsize B.Sc. Dissertation \par}
    \vspace{2cm}
    {\scshape\Large\bfseries Covariance Matrix Analysis for Optimal Portfolio Selection \par}
    \vspace{2cm}
    {\itshape\large Main supervisor: \par}
    {\large Professor \textbf{Rudy Setiono Ph.D.}\par}
    \vfill
        
    \includegraphics[width=0.5\textwidth]{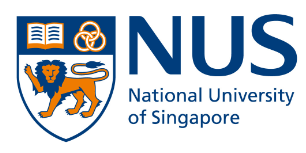}
    \vfill
    {\large Submitted by\par}
    \vspace{0.25cm}
    {\normalsize Lim Hao Shen Keith\par}
    \vspace{0.5cm}
    {\normalsize Department of Information Systems and Analytics\par}
    \vspace{0.25cm}
    {\normalsize School of Computing\par}
    \vspace{0.25cm}
    {\normalsize National University of Singapore\par}
    \vspace{1cm}
    {\normalsize 2023/2024\par}
\end{titlepage}

\begin{titlepage}
    \centering
    \vspace{1cm}
    {\normalsize B.Sc. Dissertation \par}
    \vspace{2cm}
    {\scshape\Large\bfseries Covariance Matrix Analysis for Optimal Portfolio Selection \par}
    \vspace{2cm}
        
    \includegraphics[width=0.5\textwidth]{Figures/nus-logo.png}
    \vspace{2cm}
    
    {\large Submitted by\par}
    \vspace{0.25cm}
    {\normalsize Lim Hao Shen Keith\par}
    \vspace{0.5cm}
    {\normalsize Department of Information Systems and Analytics\par}
    \vspace{0.25cm}
    {\normalsize School of Computing\par}
    \vspace{0.25cm}
    {\normalsize National University of Singapore\par}
    \vspace{0.5cm}
    {\normalsize 2023/2024\par}
    \vspace{1cm}

{\raggedright 
{\normalsize Project No: H001130\par}
{\normalsize Advisor: Professor Rudy Setiono Ph.D.\par}
\vspace{0.5cm}
{\normalsize Deliverables: \par}
{\hspace{2cm} Report: 1 Volume}

}
\end{titlepage}

\newpage

\pagenumbering{roman} 
\setcounter{page}{1} 
\begin{abstract}
    In portfolio risk minimization, the inverse covariance matrix of returns is often unknown and has to be estimated in practice. This inverse covariance matrix also prescribes the hedge trades in which a stock is hedged by all the other stocks in the portfolio. In practice with finite samples, however, multicollinearity gives rise to considerable estimation errors, making the hedge trades too unstable and unreliable for use. By adopting ideas from current methodologies in the existing literature, we propose 2 new estimators of the inverse covariance matrix, one which relies only on the $l_2$ norm while the other utilizes both the $l_1$ and $l_2$ norms. These 2 new estimators are classified as shrinkage estimators in the literature. Comparing favorably with other methods (sample-based estimation, equal-weighting, estimation based on Principal Component Analysis), a portfolio formed on the proposed estimators achieves substantial out-of-sample risk reduction and improves the out-of-sample risk-adjusted returns of the portfolio, particularly in high-dimensional settings. Furthermore, the proposed estimators can still be computed even in instances where the sample covariance matrix is ill-conditioned or singular.

    \vspace{2cm}

{\raggedright 

 Subject Descriptors:\par
\vspace{0.5cm}
\begin{tabular}{l@{\hspace{0.75em}}l}
\hspace{2cm} G.1.2 & Approximation \\
\hspace{2cm} G.1.3 & Numerical Linear Algebra \\
\hspace{2cm} G.1.6 & Optimization \\
\hspace{2cm} G.1.10 & Applications \\
\hspace{2cm} G.4 & Mathematical Software \\
\end{tabular}

\vspace{1cm}
{ Keywords:\par}
\vspace{0.5cm}
{\hspace{2cm} Data Science and Business Analytics, Portfolio Optimization,\par}
{\hspace{2cm} Quadratic Programming, Financial Analysis \par}

}
\end{abstract}
\addcontentsline{toc}{section}{Abstract}

\newpage
\section*{Acknowledgments}
\begin{doublespace} 

I wish to extend my sincere appreciation to all the individuals who offered their support and played a pivotal role throughout the duration of my Final Year Project. I am incredibly grateful to my supervisor, Professor Rudy Setiono, for the generous investment of his time and substantial expertise. His steadfast guidance and profound insights were instrumental in navigating the challenges encountered during my research. His mentorship not only facilitated my academic and intellectual growth but also played a crucial role in the realization of this thesis.

Moreover, my gratitude extends to Professor Frank Xing, my project evaluator, for his constructive criticism and feedback on my interim report and presentation. His expert advice was pivotal in refining my research direction and enhancing the quality of my work in the subsequent stages of the project. His contributions have been indispensable to the completion and success of my study.

\end{doublespace}
\addcontentsline{toc}{section}{Acknowledgements}

\newpage
\begingroup 
\color{black}
\tableofcontents
\endgroup

\newpage
\pagenumbering{arabic}
\section{Introduction}
\label{sec:introduction}
\subsection{Background}
\label{background}
Mean-variance portfolio optimization, initially introduced by \textcite{Markowitz1952}, is known for its conceptual elegance and appeal. It aims to determine optimal portfolio weights that minimize portfolio variance while targeting a specific return level. What adds to its allure is the simplicity of its analytical solution, which relies on just two key inputs: the expected mean ($\mu$) and the covariance matrix ($\Sigma$) of asset returns. Additionally, the mean-variance framework has a natural linkage to the Capital Asset Pricing Model (CAPM), a widely recognized tool in asset pricing \parencite{Ross1976}.

However, applying mean-variance optimization in practical portfolio management can be challenging. The difficulty arises from the fact that the expected mean and covariance matrix of asset returns are frequently unknown and must be estimated. Resorting to the conventional practice of substituting the sample mean and covariance matrix for the true counterparts results in the generation of extremely negative and positive portfolio weights, which are noisy and unstable over time. A substantial body of literature has documented the unsatisfactory out-of-sample performance associated with this approach (\textcite{Jobson1980}, \textcite{Jobson1981}, \textcite{BestGrauer1991}, \textcite{KanandZhou2007}, \textcite{Litterman2004}, and \textcite{Merton1980}). The lack of stability in the weights results in substantial expenses associated with portfolio rebalancing and transactions, empirically proving that the textbook estimators of the first two sample moments are inadequate in Markowitz's mean-variance portfolio optimization framework.

The estimation uncertainties come from two sources: estimation of the mean of returns and estimation of the covariance matrix of returns. It is well known that estimation errors in the mean tend to have more influence than in the covariance matrix (\textcite{BestGrauer1991}). In fact, \textcite{JagannathanMa2003} showed explicitly that the error of mean estimation is so large that nothing is lost when one ignores the mean altogether. For this reason, recent research has focused on global minimum-variance portfolios (MVP) by relying solely on the covariance estimate only, as the analytical formulation of the MVP does not require the mean as an input at all. 

Nevertheless, the MVP still suffers from errors in estimating the covariance matrix of asset returns. An extensive body of research has highlighted the inadequacy of the standard estimator, the sample covariance matrix, in addressing this issue (\textcite{Scherer2005}). One of the challenges associated with estimating covariance lies in dealing with high dimensions during portfolio construction (\textcite{fan2016}). As highlighted by \textcite{LedoitWolf2003}, \textcite{LedoitWolf2004}, and \textcite{LedoitWolf2017}, the adoption of high-dimensional scenarios in practical applications has become increasingly prevalent over the past couple of decades. Traditionally, the sample covariance matrix, denoted as $\mathbi{S}$, serves as the input for optimizing portfolio weights. However, as the portfolio dimension $p$ grows significantly relative to the sample size $n$, the matrix $\mathbi{S}$ becomes subject to notably greater estimation errors. Empirical research has revealed that these estimation errors within $\mathbi{S}$ tend to render the optimized weights less stable and less dependable, particularly when the dimension ratio $p/n$ is not small (\textcite{KanandZhou2007}), in this case, when the number of available historical return observations per stock is not much larger than the number of stocks in the portfolio. An additional source of estimation errors arises from the strong correlation among the stocks within the portfolio. \textcite{Stevens1998} presented a regression model to illustrate how multicollinearity among the stocks within a portfolio impacts the accuracy of the inverse covariance matrix estimation. To mitigate the impact of sampling errors when estimating the covariance matrix and its inverse, various methodologies have been put forth, namely to shrink the covariance matrix, and to shrink the inverse covariance matrix directly. More details on this will be elaborated in Section 2. 

On top of these methodologies, other methods have been explored to reduce the sampling errors in the covariance matrix. \textcite{JagannathanMa2003} reduced the sampling errors in the covariance matrix by constraining portfolio weights, motivated by the idea that the portfolio weights tend to take large positive and negative positions in the underlying assets. As the optimal portfolio weights take extreme values, they generate considerable turnover and are unstable. Thus, \textcite{JagannathanMa2003} constrained portfolio weights by adding no-short-sale constraints, forcing the weights to only take positive values. This implies that the MVP is exclusively long-only, indicating that investors can only purchase, not short-sell, the assets involved. Within a risk framework, this MVP could be considered less hazardous, given the inherently risky nature of short-selling.

\subsection{Objective(s) of Project}
\label{objective}
The aim of this thesis is to suggest an improvement of the estimator of the covariance matrix, or more specifically, the inverse covariance matrix ($\Sigma^{-1}$), for improving portfolio optimization and to ensure a superior out-of-sample portfolio risk reduction even when the covariance matrix and its inverse are not precisely estimated. 

The rest of this thesis is structured as follows: Section 2 provides the theoretical framework of the mean-variance optimization solution theorized by \textcite{Markowitz1952}, and examines the related work done in the area of covariance matrix estimation, particularly for portfolio optimization, and discusses some of the existing frameworks provided in recent research. Section 3 delves into the scholarly works that influenced the development of our proposed approach and meticulously outlines the mathematical framework underpinning the methodology introduced in this thesis. Section 4 then gives a detailed description of the dataset used in this thesis, and Section 5 provides a comprehensive overview of the out-of-sample portfolio performance set-up. Section 6 introduces a range of benchmark estimators utilized within the evaluation framework, alongside our proposed approach in this thesis. Section 7 outlines the performance outcomes of these estimators. Following this, Section 8 delves into discussions regarding our methodology and results, culminating in the conclusive insights in Section 9, and suggestions for future enhancements in Section 10.

\newpage
\section{Literature Review}
\label{literature review}
In this section, several of the most recent and notable advancements made in the field of portfolio optimization are discussed and highlighted, going back to the formation of the mean-variance portfolio optimization by \textcite{Markowitz1952}. 

\subsection{Mean-Variance and Global Minimum-Variance portfolios}
\label{sec:sec21}
To better understand the motivation and purpose of this thesis, it is imperative to understand how the mean-variance optimization works. Let $r_t$ denote the vector of asset returns on $p$ stocks with covariance matrix $\Sigma$ at time $t$. In the traditional mean-variance framework, the optimal portfolio weights are obtained by minimizing the overall risk of the portfolio under certain constraints: 
\begin{equation}
\begin{aligned}
w^{\ast} = argmin_{w} \quad w^{\top}\Sigma w \\
\textrm{s.t.} \quad w^{\top}e = 1\\
w^{\top}\mu = r\\
\end{aligned}
\end{equation}
where $w$ is the column vector of length $p$ of portfolio weights, $e$ is the vector of length $p$ with all entries equal to 1, $\mu$ is a column vector of length $p$ denoting expected return of each asset, and $r$ is the expected return of the portfolio wanted. Note that the objective function $w^{\top}\Sigma w$ is a convex function of $w$, and the constraints $w^{\top}e = 1$ and $w^{\top}\mu = r$ are linear in $w$. Hence the optimization problem can be transformed into the equivalent Lagrangian form as 

\begin{equation}
argmin_{w} \quad w^{\top}\Sigma w - \lambda_{1}(w^{\top}e - 1) - \lambda_{2}(w^{\top}\mu - r)
\end{equation}
where $\lambda_{1}$ and $\lambda_{2}$ are the Lagrangian multipliers

By differentiating equation (2) with respect to $w$, $\lambda_{1}$, and $\lambda_{2}$, and solving it, we can derive the analytical solution to be:

\begin{equation}
w^{\ast} = \frac{c-br}{ac-b^2}\Sigma^{-1}e + \frac{ar-b}{ac-b^2}\Sigma^{-1}\mu
\end{equation}
where $a = e^{\top}\Sigma^{-1}e$, $b = e^{\top}\Sigma^{-1}\mu$, and $c = \mu^{\top}\Sigma^{-1}\mu$. Refer to Appendix B for a more comprehensive derivation.

As we can see, the original mean-variance framework relies on the two input parameters $\mu$ and $\Sigma$. As mentioned in Section 1, extensive literature has studied the global minimum-variance portfolios instead, as it removes the need for the mean as an input. To find the MVP, the only constraint required is $w^{\top}e = 1$, as follows:
\begin{equation}
\begin{aligned}
w^{\ast} = argmin_{w} \quad w^{\top}\Sigma w \\
\textrm{s.t.} \quad w^{\top}e = 1\\
\end{aligned}
\end{equation}
The explicit solution to the MVP is simply given by
\begin{equation}
w^{\ast} = \frac{\Sigma^{-1}e}{e^{\top}\Sigma^{-1}e}
\end{equation}
which only depends on the covariance matrix $\Sigma$, or more specifically the inverse covariance matrix $\Sigma^{-1}$. As can be seen from the simple analytical formulation of the MVP which only requires the inverse covariance matrix as the input, the inverse covariance matrix is thus widely known to be the "optimizer" in the MVP framework.

As such, extensive literature has explored various methodologies to improve the estimator of the covariance matrix and the inverse covariance matrix. From here on, we will denote the inverse covariance matrix as the precision matrix. 

\subsection{Covariance matrix estimation}
In order to reduce sampling errors in estimating the covariance matrix, earlier research has explored shrinking the covariance matrix. A typical way to shrink the covariance matrix is based on Bayes-Stein shrinkage estimators (\textcite{Jorion1986}), which introduce a balance between variance and bias by shrinking the sample covariance matrix toward a prior matrix. This prior matrix is a prespecified common value proposed to shrink the covariance matrix towards. \textcite{LedoitWolf2003} and \textcite{LedoitWolf2004} further developed the idea by imposing structure on the estimator. \textcite{LedoitWolf2003} do so by shrinking the sample covariance matrix towards the single-index covariance matrix by \textcite{Sharpe1963}, while \textcite{LedoitWolf2004} performed the shrinkage towards the identity matrix. The single-index covariance matrix is the covariance matrix of the single-index model developed by \textcite{Sharpe1963}, where it is assumed that there is only one macroeconomic factor that causes the systematic risk affecting all stock returns and this factor can be represented by the rate of return on a market index. Both work on the fundamental principle of statistical decision theory that there exists an interior optimum in the trade-off between bias and estimation error, and one way of attaining this optimal trade-off is simply to take a properly weighted average of the biased (single-index model or identity matrix) and unbiased (sample covariance matrix) estimators, which results in a shrinking of the unbiased estimator full of estimation error towards a fixed target represented by the biased estimator. More specifically, \textcite{LedoitWolf2004} derived the formula for the shrinkage estimator:
\begin{equation}
\Sigma_{shrunk} = (1-\alpha)\Sigma + \alpha \overline{\sigma^{2}}I
\end{equation}
where $\alpha$ denotes the shrinkage intensity, with $0 < \alpha < 1$, and $\overline{\sigma^{2}}$ is the average of the diagonal elements of the sample covariance matrix, which gives the average variance of the underlying assets.

\subsection{Precision matrix estimation}

Beyond applying shrinkage techniques to the sample covariance matrix, there is also a practice of applying shrinkage to the precision matrix. Some modern methodologies are centered around this concept. This may be because the precision matrix exerts a more direct influence on portfolio weights compared to the covariance matrix, as the covariance matrix has to be further inverted in the MVP framework. With an estimated precision matrix in hand, portfolio weights can be directly derived, eliminating the necessity of computing the matrix inverse. The estimation of high-dimensional covariance matrices and their inverses has emerged as a substantial and dynamic research area in recent years. Recent works have shown two important avenues for estimating a high-dimensional precision matrix: structure-based estimation and structure-free estimation. Structure-based estimation makes the problem more amenable by assuming additional structure in the precision matrix, while structure-free estimation requires no prespecification of a structured form of the covariance matrix and its inverse. As our proposed methodology is extended from this, we will be going deeper in detail in this section.

\subsubsection{Structure-based estimation}

The foundation for this approach stems from the insight of \textcite{Stevens1998}, where the precision matrix $\Psi = \Sigma^{-1}$ unveils hedging strategies among stocks. To be specific, the $i^{th}$ row (or column) of $\Psi$ reflects the minimum-variance hedge portfolio for the $i^{th}$ stock. This hedge portfolio involves taking a long position in the $i^{th}$ stock and a short position in the "tracking portfolio" of the remaining $N - 1$ stocks to hedge against the $i^{th}$ stock's movements. The off-diagonal elements of the precision matrix represent scaled coefficients obtained through Ordinary Least Squares (OLS) regression, showcasing the relationships of each stock with the other $N - 1$ stocks in the portfolio, while the diagonal elements represent the unhedgeable risk of the $i^{th}$ stock. 

However, in a large portfolio, it is not necessarily ideal to hedge each specific stock with the other $N - 1$ stocks in small samples. Multicollinearity introduces substantial estimation errors in these coefficients. To bolster the robustness of optimization, multiple studies have explored strategies akin to those used to mitigate multicollinearity issues in regression. This involves applying regularization to shrink coefficients within each hedge regression toward zero, thereby reducing extreme hedge positions and eliminating redundant stocks within each hedge portfolio. The outcome is a sparse mean-variance optimizer.

Following \textcite{Stevens1998} framework, one can apply the least absolute shrinkage and selection operator (Lasso) (\textcite{Tibshirani1996}) to each hedge regression to estimate each row (or column) of the precision matrix to achieve shrinkage and variable selection. This was in fact explored by \textcite{Meinshausen2006}, which although showed that it correctly estimates the nonzero elements, it does not restrict the precision matrix to be positive definite or symmetric. Hence the $N$ hedge regressions have to be estimated jointly as a group instead. \textcite{Yuan2007}, as well as \textcite{Rothman2008}, estimated $\Psi$ through Quasi-Maximum Likelihood (QML) while imposing constraints on the sum of the absolute values of its off-diagonal elements. On the other hand, \textcite{Goto2015} employed the Graphical Lasso (glasso) algorithm introduced by \textcite{Friedman2008} to address the estimation challenge. More mathematical details will be introduced in the subsequent sections.

\subsubsection{Structure-free estimation}

When there are no a priori beliefs about the matrix structure, structure-free estimation is the most general and neutral approach (\textcite{shi_shu_yang_he_2020}). The key idea of structure-free estimation is to estimate the covariance matrix and its inverse by reducing the errors in its eigenvalues. It is worth noting that the eigenvalues and eigenvectors of the precision matrix provide a unique means of ascertaining the matrix's characteristics through spectral decomposition. \textcite{Marcenko1967} showed that when the dimension of the matrix is not small relative to the sample size, the eigenvalues of the sample covariance matrix and its inverse will be more dispersed than the true ones. Moreover, \textcite{shi_shu_yang_he_2020} showed that sampling errors in the precision matrix are mainly dominated by errors in the eigenvalues, while the effect of sampling errors in eigenvectors is only marginal compared to that of the eigenvalues. The condition-number-regularized estimator proposed by \textcite{conditionnumber2013} bounds the condition number with the goal of obtaining a well-conditioned estimator by winsorizing the sample eigenvalues at the lower and upper end of the spectrum. \textcite{shi_shu_yang_he_2020} introduced a computationally efficient estimator of the precision matrix by alleviating overdispersion of the sample eigenvalues through a constraint on the Schatten $q$-norm of the precision matrix. The Schatten $q$-norm represents a family of norms, where it is computed as the $q$-norm of the vector of singular values, as shown below:
\begin{equation}
||\Psi||_{q} = (\sum_{i=1}^{p}m_{i}^{q})^{\frac{1}{q}}
\end{equation}
where $m_{i}$ is the $i^{th}$ largest eigenvalue of matrix $\Psi$.

\newpage
\section{Proposed Methods}
The proposed methods that we will put forth in this paper are an extension of the application of the glasso algorithm to portfolio optimization as introduced in \textcite{Goto2015}. Specifically, instead of imposing a penalty on the $l_1$ norm of the off-diagonal elements of the precision matrix, our proposed methods involve imposing a penalty on the $l_2$ norm, and a linear combination of the $l_1$ norm and the $l_2$ norm, of the off-diagonal elements. Analogously, this would be converting a Lasso regression to a Ridge regression and an Elastic Net regression in the context of OLS regression. 

\subsection{Relation to the Literature}
As described in section 2.3.1, \textcite{Stevens1998} provided the insight that the precision matrix, $\Psi$, unveils hedging strategies among stocks, where each row (or column) of $\Psi$ reflects the minimum-variance hedge portfolio for the $i^{th}$ stock, which involves taking a long position in the $i^{th}$ stock and a short position in the "tracking portfolio" of the remaining $N-1$ stocks to hedge against the $i^{th}$ stock's movements. 
The other $N - 1$ stocks are called the "tracking portfolio" as they track the $i^{th}$ stock return to minimize the variance of the error term in the regression. Mathematically, each row (or column) of $\Psi$ consists of the regression:

\begin{equation}
r_{i,t} = \alpha_i + \sum_{\substack{j=1, j \neq i}}^{N} \beta_{i|j}r_{j,t} + \varepsilon_{i,t}
\end{equation}
where $r_{i,t}$ denotes the $i^{th}$ stock return in period $t$; $\varepsilon_{i,t}$ is the unhedgeable component of $r_{i,t}$, whose variance is denoted by $v_i = var(\varepsilon_{i,t})$; and $v_i$ is a measure of the $i^{th}$ stock's unhedgeable risk. The objective of each hedge regression is to minimize $v_i$ and hence we can view the regression as an OLS estimation problem. When $\beta_{i|j}$ is different from 0, it implies that the $j^{th}$ stock provides a marginal contribution to the hedge of the $i^{th}$ stock beyond the contributions of the other $N-2$ stocks in the portfolio.  

Extending this concept to the precision matrix, the off-diagonal elements of the precision matrix represent scaled coefficients obtained through OLS regression, showcasing the relationships of each stock with the other
$N-1$ stocks in the portfolio, while the diagonal elements represent the unhedgeable risk of the $i^{th}$ stock.

In fact, \textcite{Stevens1998} establishes the following identity between $\Psi$ and hedge regression:

\begin{equation}
\psi_{ij} = 
\begin{cases}
\mbox{\LARGE $-\frac{\beta_{i|j}}{v_i}$} & \mbox{if $i \neq j$} \\
\mbox{\LARGE $\frac{1}{v_i}$} & \mbox{if $i = j$}
\end{cases}
\end{equation}
We can view $\psi_{ij}$ as a measure of marginal hedgeability between the $i^{th}$ and $j^{th}$ stocks conditional on all other stocks in the portfolio, where $\psi_{ij}$ denotes the $(i,j)^{th}$ element of $\Psi$.

Hence, each row in $\Psi$ looks like:
\begin{equation}
\Psi (i, \cdot) = \frac{1}{v_i} 
\begin{bmatrix}
-\beta_{i|1}, & \cdots, & -\beta_{i|i-1}, & 1, & -\beta_{i|i+1}, & \cdots, & -\beta_{i|N},
\end{bmatrix}.
\end{equation}
This equation can be regarded as a vector of asset holdings in the $i^{th}$ stock's hedge portfolio. It involves a unit long in the $i^{th}$ stock and a short position of the hedge portfolio constructed from regression (8). These equations are useful for understanding the sources of potential large estimation errors in mean-variance portfolio optimization. 

As seen in equation (8), each hedge regression contains a constant and $N-1$ stock returns as regressors. However, in practice, many of these stock returns are highly correlated. Furthermore, the number of available historical returns to estimate the regression is not much larger than the number of stocks. In other words, the number of observations is not enough compared to the dimension of the data. This would subject the regressions to multicollinearity, which in turn results in the estimated coefficients ($\hat{\beta}$) having large estimation errors. 

By framing the rows (or columns) of $\Psi$ to be regression estimations, the classic methods used in regressions to improve the estimation can be applied to this problem. In particular, regularization can be applied to the regressions to reduce multicollinearity between the regressors and to improve the estimation of the coefficients. For example, the Lasso regression introduced by \textcite{Tibshirani1996} improves on the basic regression model by introducing the $l_1$ norm as a regularization term to the regression. The $l_1$ norm of the regression is the sum of the absolute values of the coefficients of the regression. Specifically, we consider the Lasso estimation problem for the hedge regression in equation (8):

\begin{equation}
\hat{\beta}_{i|j}^{\text{lasso}} = \arg\min_{\beta} \left\{ \sum_{t=1}^{T} \left( r_{i,t} - \sum_{\substack{j=1, j \neq i}}^{N} \beta_{i|j}r_{j,t} \right)^2 + \gamma \sum_{\substack{j=1, j \neq i}}^{N} |\beta_{i|j}| \right\}.
\end{equation}
If the regressors were orthonormal, the Lasso coefficient solving the above equation would have the relationship with the OLS coefficient $\hat{\beta}_{i|j}^{OLS}$:

\begin{equation}
\hat{\beta}_{i|j}^{\text{lasso}} = \text{sign}\left(\hat{\beta}_{i|j}^{\text{OLS}}\right) \left( \left|\hat{\beta}_{i|j}^{\text{OLS}}\right| - \gamma \right)_+ , \quad j = 1, \ldots, N, \quad j \neq i,
\end{equation}
where $(x)_+$ = $max(x, 0)$ and the penalty parameter $\gamma$ becomes the soft threshold. The OLS coefficients $\hat{\beta}_{i|j}^{OLS}$ can be estimated through the analytical formula of the standard regression, given by: 

\begin{equation}
\hat{\beta}^{OLS} = (\mathbf{X}^T\mathbf{X})^{-1}\mathbf{X}^T\mathbf{y}
\end{equation}
where $\mathbf{X}$ is the design matrix of the data points, and $\mathbf{y}$ is the vector of response.

When the magnitude of an OLS point estimate is below the soft threshold, Lasso solution $\hat{\beta}_{i|j}$ is set to 0, while those that are above the threshold are shrunk by the size of the soft threshold. In this way, Lasso achieves both shrinkage and sparsity, which in the context of the hedge regression, means the selection of a subset of stocks in each hedge portfolio. 

In practice, the regressors are hardly orthonormal, as they are usually correlated with each other, given that we are working with stock data. As such, the $\hat{\beta}_{i|j}^{\text{lasso}}$ has to be estimated iteratively.
\textcite{Friedman2007} showed that starting from $\hat{\beta}_{i|j}^{OLS}$, repeated iteration of the following yields the Lasso estimate:

\begin{equation}
\hat{\beta}_{i|j}^{(\gamma)} = \mathbf{S} \left( \hat{\beta}_{i|j}^{(\gamma)} + \sum_{t=1}^{T} \left( r_{j,t} \times \left( r_{i,t} - \sum_{\substack{k=1, k \neq i}}^{N} \hat{\beta}_{i|k}^{(\gamma)} r_{k,t} \right)\right), \gamma \right) ; \quad j = 1, \ldots, N, \quad j \neq i,
\end{equation}
where $\mathbf{S}(b, \gamma)$ = $sign(b)(|b| - \gamma)_+$ is the soft thresholding operator. The mathematical framework and introduction above form the basis for the subsequent introduction of the glasso framework by \textcite{Goto2015}. 

Although applying the Lasso to each hedge regression seems promising, \textcite{Meinshausen2006} showed that the resulting precision matrix that is formed may not necessarily be positive definite nor symmetric, both of which are important characteristics of the precision matrix. As such, the Lasso should not be applied to each row independently but should be done jointly, and \textcite{Yuan2007}, and \textcite{Friedman2008} introduced an approach to estimate all elements of the precision matrix once for all, which involves solving the QML estimation with a penalty on the $l_1$ norm of its off-diagonal elements:

\begin{equation}
\max_{\Psi = [\psi_{ij}]} \frac{T}{2} \ln(\det(\Psi)) - \frac{T}{2} \text{trace}(\hat{S}\Psi) - \rho \sum_{\substack{i=1, i\neq j}}^{N} \sum_{\substack{j=1, j\neq i}}^{N} |\psi_{ij}|,
\end{equation}
where det and trace indicate matrix determinant and trace respectively, and $\hat{S}$ is the sample covariance matrix. The regularization parameter $\rho$ denotes the penalty on the $l_1$ norm of the off-diagonal elements of $\Psi$, given as $\sum_{\substack{i=1, i\neq j}}^{N} \sum_{\substack{j=1, j\neq i}}^{N} |\psi_{ij}|$, and it should be noted that $\rho \geq 0$, with $\rho = 0$ indicating no penalty at all, equivalent to the unconstrained problem. Furthermore, it should be noted that the equation $\ln(\det(\Psi)) - \text{trace}(\hat{S}\Psi)$ denotes the log-likelihood function, which is always negative except when at optimality where it gives 0. With the inclusion of the penalty of the norm, we are maximizing both the log-likelihood function and the negative of the penalty term, hence this is a maximization problem. \textcite{Goto2015} utilized the glasso algorithm by \textcite{Friedman2008} to solve this QML estimation problem, which was demonstrated to be equivalent to an $N$-coupled Lasso problem, and showed that it lowers the estimation error and the condition numbers of the estimated precision matrix compared to the sample estimator. The glasso algorithm works by sweeping over each row in the precision matrix to solve individual Lasso problems, while holding all the others fixed, which then proceeds to the next step until all coefficients converge.

\subsection{Proposed Methods}
\subsubsection{Ridge}
In this thesis, we suggest modifying the Quasi-Maximum Likelihood (QML) estimation approach by utilizing the $l_2$ norm as a substitute for the $l_1$ norm. This concept is numerically articulated in equation (8), which interprets each row (or column) of the precision matrix in the context of a regression problem. Furthermore, as shown in equation (11), the introduction of regularization to each regression can be applied to improve the estimation. We assert that a widely recognized variant of regression, Ridge regression, is also suitable for application within this framework. 

Ridge regression, introduced by \textcite{Hoerl1970}, operates by adding a degree of bias to the regression estimates, which results in significant reductions in variance. Specifically, it introduces a penalty term equal to the square of the magnitude of the coefficients to the loss function. This method, known in mathematical terms as $l_2$ regularization, effectively shrinks the coefficient estimates towards zero. The primary advantage of this shrinkage is that it reduces model complexity and prevents overfitting, which is particularly useful when dealing with multicollinearity or when the number of predictors exceeds the number of observations. By incorporating this regularization parameter, Ridge regression improves the model's predictive performance and robustness, especially in scenarios where the predictors are highly correlated. On top of that, the $l_2$ norm penalty tends to penalize the extremes of the weights, resulting in a group of weights that are more evenly distributed. 

By incorporating the $l_2$ norm into the regression equation in (8), the Ridge regression formulation can be described as shown:

\begin{equation}
\hat{\beta}_{i|j}^{\text{ridge}} = \arg\min_{\beta} \left\{ \sum_{t=1}^{T} \left( r_{i,t} - \sum_{\substack{j=1, j \neq i}}^{N} \beta_{i|j}r_{j,t} \right)^2 + \gamma \sum_{\substack{j=1, j \neq i}}^{N} \beta_{i|j}^2 \right\}.
\end{equation}
The key difference from the Lasso regression is the penalty norm to the objective function. The $\gamma$ parameter controls the magnitude of the penalty to the regression problem, similar to that of the Lasso. Following the QML estimation which estimates the regressions jointly, the $l_2$ norm can be used to replace the $l_1$ norm as well, as shown below:

\begin{equation}
\max_{\Psi = [\psi_{ij}]} \frac{T}{2} \ln(\det(\Psi)) - \frac{T}{2} \text{trace}(\hat{S}\Psi) - \rho \sum_{\substack{i=1, i\neq j}}^{N} \sum_{\substack{j=1, j\neq i}}^{N} \psi_{ij}^2
\end{equation}
The objective of applying the $l_2$ norm to the non-diagonal components of the estimated precision matrix is to achieve a reduction in the matrix's extreme figures through shrinkage. While the application of the $l_1$ norm brings about both sparsity and shrinkage in the precision matrix, utilizing the $l_2$ norm potentially offers superior shrinkage capabilities for those extreme values.

\subsubsection{Elastic Net}
On top of utilizing the $l_2$ norm as a substitute for the $l_1$ norm, we also propose investigating the Elastic Net as a method of regularization. \textcite{hastie2005} introduced the Elastic Net as a regularization technique that blends properties of both Lasso ($l_1$ regularization) and Ridge ($l_2$ regularization) regression methods. It aims to minimize the loss function by incorporating both $l_1$ and $l_2$ penalties, which helps to handle issues when there are correlations between the variables, as well as with feature selection. In situations where Lasso might select one variable from a group of correlated variables, the Elastic Net is more likely to select the group as a whole, which can often result in better model performance. In fact, \textcite{hastie2005} explained that if there is a group of variables among which the pairwise correlations are very high, then the Lasso tends to select only one variable from the group and does not care which one is selected. \textcite{Tibshirani1996} also showed that if there are high correlations between predictors, it has been empirically observed that the prediction performance of the Lasso is dominated by Ridge regression. 

The key advantage of Elastic Net is that it allows for the selection of multiple correlated features and provides a more balanced approach to regularization, combining feature elimination from Lasso and feature shrinkage from Ridge. This can be particularly beneficial in cases where there are fewer observations than predictors, or when several predictors are highly correlated with each other.

By incorporating the Elastic Net into the regression equation in (8), the Elastic Net regression formulation can be described as shown:

\begin{equation}
\hat{\beta}_{i|j}^{elasticnet} = \arg\min_{\beta} \left\{ \sum_{t=1}^{T} \left( r_{i,t} - \sum_{\substack{j=1, j \neq i}}^{N} \beta_{i|j} r_{j,t} \right)^2 + \gamma \left((1-\alpha) \sum_{\substack{j=1, j \neq i}}^{N} |\beta_{i|j}| + \alpha \sum_{\substack{j=1, j \neq i}}^{N} \beta_{i|j}^2 \right) \right\}
\end{equation}
The Elastic Net has two tuning parameters: $\alpha$, which controls the balance between $l_1$ and $l_2$ regularization, and $\gamma$, the penalty term that affects the amount of shrinkage applied to the coefficients, similar to that of the Lasso and Ridge regressions shown above. Hence it can be seen that the Elastic Net norm is simply a linear combination of the $l_1$ and $l_2$ norms. By adjusting these parameters, one can calibrate the model to find the right balance between the complexity and the generalizability of the model. The optimal values for these parameters are typically determined through cross-validation.

Incorporating this into the QML estimation problem as in equation (17), the estimation of the precision matrix using the Elastic Net can be shown below:
\begin{equation}
\max_{\Psi = [\psi_{ij}]} \frac{T}{2} \ln(\det(\Psi)) - \frac{T}{2} \text{trace}(\hat{S}\Psi) - \rho \left( (1-\alpha)\sum_{\substack{i=1, i\neq j}}^{N} \sum_{\substack{j=1, j\neq i}}^{N} |\psi_{ij}| + \alpha\sum_{\substack{i=1, i\neq j}}^{N} \sum_{\substack{j=1, j\neq i}}^{N} \psi_{ij}^2 \right)
\end{equation}
Utilizing the Elastic Net within the QML estimation integrates the strengths of both regularization techniques, refining the estimation by overcoming the constraints that arise when each norm is applied separately. This method aims to estimate a more robust precision matrix by harmonizing the regularization effects of both norms.

Both our proposed methods as well as the glasso estimator proposed by \textcite{Goto2015} belong to a wide class of shrinkage estimators. As such, it shares the same objective as a few existing approaches in that it shrinks the estimator from the unbiased estimator, the sample estimator, in the direction that reduces estimation errors (\textcite{LedoitWolf2003}, \textcite{LedoitWolf2004}, \textcite{LedoitWolf2017}).

It is important to mention that while the study by \textcite{Goto2015} employed the glasso algorithm, originally developed by \textcite{Friedman2007}, to address the QML estimation problem, in our approach, the QML estimation is tackled independently through the application of convex optimization. This involves directly setting up and solving for the objective function and constraints within the optimization framework.

\newpage
\section{Data}
\label{data}

\subsection{Data Description}
For the evaluation of the estimators, we use some public datasets that are widely used in the current literature. These datasets consist of well-diversified portfolios. The datasets are extracted from the website of Ken French (\url{http://mba.tuck.dartmouth.edu/pages/faculty/ken.french/data_library.html#Benchmarks}), including 17-, 30-, and 49-industry portfolios (17Ind, 30Ind, 49Ind, respectively); 100 portfolios formed on size and book-to-market ratio (100FF); and the combination of the 100FF and the 32 portfolios formed on size, book-to-market, and operating profitability (132S). 

The datasets for each portfolio were analyzed for monthly average value-weighted returns within the timeframe of July 1973 to December 2015, following the methodology of \textcite{shi_shu_yang_he_2020}. This period encompasses a total of 510 observations. From these datasets, various metrics were calculated, including the dataset dimensionality, denoted as $p$, the ratio of dimensionality to the sample size $p/n$, alongside the mean absolute correlation and the maximum correlation observed within the data. The results are shown in Table 1.

\begin{table}[htbp]
\begin{center}
\begin{tabular}{||c c c c c||} 
 \hline
 Data & $p$ & $p/n$ & Maximum Correlation & Mean Absolute Correlation \\ [1ex]
 \hline
 17Ind & 17 & 0.03 & 0.87 & 0.59 \\ [1ex]
 \hline
 30Ind & 30 & 0.06 & 0.86 & 0.58 \\[1ex]
 \hline
 49Ind & 49 & 0.10 & 0.86 & 0.55 \\[1ex]
 \hline
 100FF & 100 & 0.20 & 0.96 & 0.69 \\[1ex]
 \hline
 132S & 132 & 0.26 & 0.96 & 0.70 \\  [1ex]
 \hline
\end{tabular}
\end{center}
\caption{Data Description}
\label{table:1}
\end{table}

An increase in the dataset's dimensionality typically accentuates the correlations among the variables. This phenomenon is evident in the data from the 132S dataset, which recorded a near-perfect correlation coefficient of 0.96. Such a high correlation implies that within the dataset, there exist portfolios whose performance is nearly identical, suggesting a strong linear relationship between them. This observation empirically shows the existence of multicollinearity among the variables, in this case, the portfolios in the dataset. Nevertheless, this is unsurprising, as they are influenced by a series of common factors such as underlying economic conditions, business cycle fluctuations, and technological advancements in supply chain management. These elements have a widespread impact on the financial markets, leading to interconnected movements in asset prices (Boldrin et al. (2001), Beaubrun-Diant and Tripier (2005), Menzly and Ozbas (2010)).

\begin{figure}[h]
\centering
\includegraphics[width=1\textwidth]{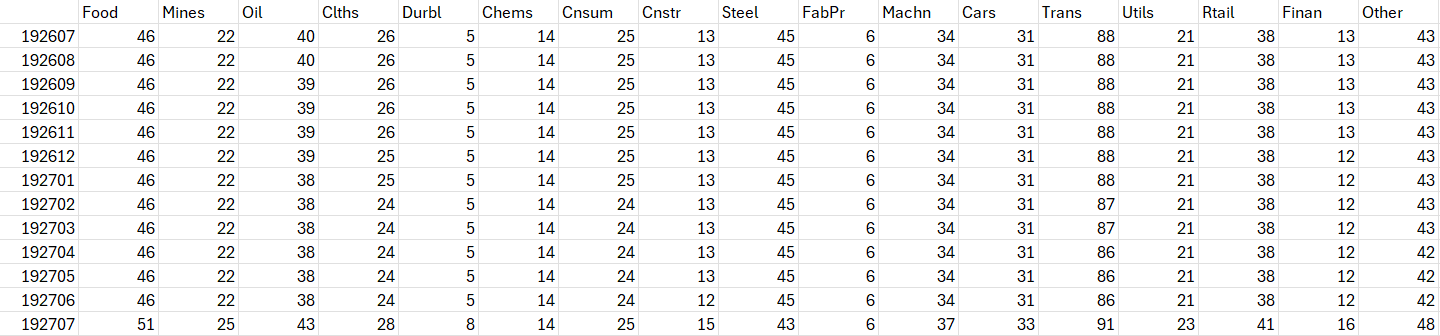}
\caption{Number of companies for each industry in 17Ind}
\label{fig:your-label}
\end{figure}

Figure 1 presents a breakdown of the number of firms within each industry in the 17Ind dataset, spanning from July 1926 through July 1927. This depiction reveals a generally stable count of companies across industries throughout the observed period. However, minor fluctuations are noted, such as an increment in the Food industry's firm count, rising from 46 in June 1927 to 51 by July 1927. Monthly returns for each industry portfolio, represented in the 17Ind dataset by a column, are calculated as the value-weighted mean of monthly returns from individual companies within that industry. Detailed methodologies for calculating these industry-specific portfolios and the monthly returns of the constituent assets are not disclosed on the associated website.

Despite being the dataset with the least dimensions, the 17Ind dataset still registers a high maximum correlation of 0.87, underlining significant correlation levels even when segregating the data by industry groups. Due to the large size of the other datasets, they are not displayed here; only the 17Ind dataset with its number of companies is shown for illustrative purposes. However, it is important to note that the remaining datasets also feature company counts which remain relatively stable over the years, without notable fluctuations.

\subsection{Exploratory Data Analysis}
The work presented in this section aims to provide an exploratory data analysis of the datasets used in this thesis, with the insights incorporated in the subsequent data preprocessing step to enhance the performance of the estimators used in this study.

\subsubsection{Missing Data}
On preliminary analysis of the dataset, we discovered that missing data were indicated by -99.99 or -999 in the datasets. As such, we had to convert these extreme and unlikely values into null first and count the number of occurrences. Thereafter, we filled the null values using the forward-filling method, as using the standard data imputation methods like filling with the mean or median would induce look-ahead bias in time series data. 
Look-ahead bias occurs in the analysis of historical data when a model or algorithm inadvertently uses information that would not have been available or known during the period being analyzed. For example, if we impute the null values using the mean or median of the column, it would use future data that would not have been available at that specific time point, because the computation of the mean or median would utilize the entire column. To give a more specific example, if we have a missing value in June 2000 for the Food industry portfolio in the 17Ind dataset, filling it using the mean or the median of the column would utilize the full data from July 1973 to December 2015 in the computation of the mean or median, but as of June 2000, any data point beyond that is meant to be unavailable at that point in time. Forward-filling works by replacing missing values with the most recent valid (non-missing) value that occurs before them in the dataset. In other words, forward-filling propagates the last known value forward in the time series or along the rows. This would overcome the look-ahead bias that would exist in standard data imputation methods like imputing by the mean or median. The table below shows the number of null values for each dataset.

\begin{table}[htbp]
\begin{center}
\begin{tabular}{||c c||} 
 \hline
 Data & No. of Nulls \\ [1ex]
 \hline
 17Ind & 0 \\ [1ex]
 \hline
 30Ind & 0 \\ [1ex]
 \hline
 49Ind & 0 \\ [1ex]
 \hline
 100FF & 161 \\ [1ex]
 \hline
 132S & 161 \\  [1ex]
 \hline
\end{tabular}
\end{center}
\caption{Null Counts per Dataset}
\label{table:1}
\end{table}

From Table 2, it is clear that the 100FF and 132S datasets contain the most nulls, and since the 132S dataset consists of the combination of 100FF dataset with the 32 portfolios formed on size, book-to-market, and operating profitability, it is clear that the nulls came from the 100FF dataset. 

\subsubsection{Returns}
On analysing the data, we are interested in visualizing how each asset performs over time. It is worth noting that the datasets used consists of well-diversified portfolios, which means that each asset, or column, in the dataset is a diversified portfolio on its own. To visualize how these portfolios perform over time, we chose to plot the returns of the 17Ind dataset against the time period of interest, as seen in Figure 2. The returns data against time for the rest of the datasets were inspected and shown in Appendix A.
\begin{figure}[htbp]
\begin{center}
\includegraphics[width=17cm]{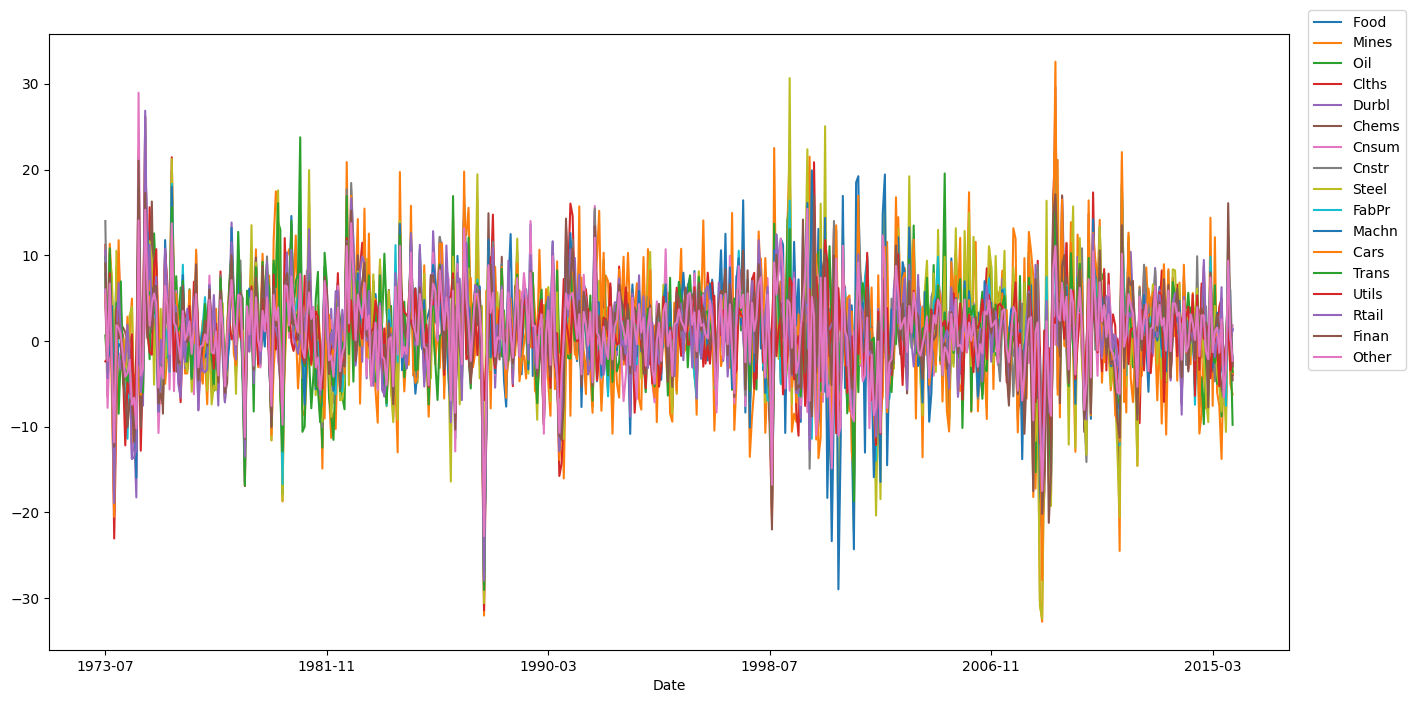}
\caption{Plot of returns against time}
\end{center}
\end{figure}

At first glance, it can be seen that the returns data looks to be exhibiting stationarity properties, in that there does not seem to be any obvious trends or seasonality across time. \textcite{Hsu1984} also stated that there are weak correlations between stock returns in consecutive time periods, and that most research work implicitly relies on the stationarity assumption. To prove for stationarity beyond simply visualizing the time series, the Augmented Dickey-Fuller (ADF) test introduced by \textcite{dickeyfuller} can be used to test if a time series is stationary. The ADF test is a statistical test that tests the null hypothesis that a unit root is present, which would indicate non-stationarity. A low p-value indicates strong evidence against the null hypothesis of the presence of a unit root, suggesting that the time series is stationary. Conversely, a high p-value suggests that the time series has a unit root and is non-stationary. For this test, we would use a threshold of 0.05 for the p-value.

\begin{table}[htbp]
\begin{center}
\begin{tabular}{||c c c||} 
 \hline
 Industry & ADF Statistic & p-value \\ [1ex] 
 \hline
 Food & -9.70 & 1.06e-16\\ [1ex]
 \hline
 Mines & -21.85 & 0.00 \\[1ex]
 \hline
 Oil & -22.99 & 0.00 \\[1ex]
 \hline
 Clths & -15.78 & 1.13e-28 \\[1ex]
 \hline
 Durbl & -20.15 & 0.00 \\  [1ex]
 \hline
 Chems & -21.96 & 0.00 \\  [1ex]
 \hline
 Cnsum & -7.07 & 4.79e-10 \\  [1ex]
 \hline
 Cnstr & -10.64 & 5.09e-19 \\  [1ex]
 \hline
 Steel & -21.77 & 0.00 \\  [1ex]
 \hline
 FabPr & -16.18 & 4.24e-29 \\ [1ex]
 \hline
 Machn & -21.62 & 0.00 \\  [1ex]
 \hline
 Cars & -20.42 & 0.00 \\ [1ex]
 \hline
 Trans & -16.33 & 3.03e-29 \\ [1ex]
 \hline
 Utils & -8.72 & 3.27e-14\\ [1ex]
 \hline
 Rtail & -10.80 & 1.95e-19 \\  [1ex]
 \hline
 Finan & -9.51 & 3.17e-16 \\  [1ex]
 \hline
 Other & -21.19 & 0.00 \\ [1ex]
 \hline
\end{tabular}
\end{center}
\caption{ADF Statistic and p-values for each Industry portfolio in 17Ind}
\label{table:1}
\end{table}

Table 3 shows the test statistic as well as the p-value of the ADF test applied to each column of the 17Ind dataset. Since each column corresponds to the time series of an industry portfolio, the ADF test is applied individually to each column of the dataset. The ADF test shows that all the portfolios are stationary, as they have p-values that are either 0 or extremely small, all of which are $<$ 0.05. Thus there does not seem to be any peculiar problems with the returns in the dataset.

\subsubsection{Sharpe and Variance}
To gain deeper insights into the dataset, calculations of the Sharpe ratio and variance for each industry portfolio within the 17Ind dataset are undertaken for the entire period. This analysis not only helps in assessing the risk-adjusted returns of each industry portfolio but also in understanding the volatility or risk inherent in each industry over time. The Sharpe ratio is computed by taking the mean returns divided by the standard deviation of the returns for the entire sample:

\begin{equation}
\hat{SR} = \frac{\hat{\mu}}{\hat{\sigma}}
\end{equation}
where $\hat{\mu}$ and $\hat{\sigma}$ denotes the sample mean and sample standard deviation respectively. The Sharpe ratio and the variance of each portfolio are shown in Table 4.

\begin{table}[htbp]
\begin{center}
\begin{tabular}{||c c c||} 
 \hline
 Industry & Sharpe Ratio & Variance \\ [0.5ex] 
 \hline
 Food & 0.259 & 20.09\\ [0.5ex]
 \hline
 Mines & 0.119 & 61.87 \\ [0.5ex]
 \hline
 Oil & 0.188 & 32.19 \\[0.5ex]
 \hline
 Clths & 0.188 & 39.75 \\[0.5ex]
 \hline
 Durbl & 0.124 & 33.54 \\  [0.5ex]
 \hline
 Chems & 0.175 & 34.94 \\  [0.5ex]
 \hline
 Cnsum & 0.231 & 22.55 \\  [0.5ex]
 \hline
 Cnstr & 0.191 & 37.93 \\  [0.5ex]
 \hline
 Steel & 0.101 & 60.08 \\  [0.5ex]
 \hline
 FabPr & 0.184 & 30.53 \\ [0.5ex]
 \hline
 Machn & 0.144 & 45.80 \\  [0.5ex]
 \hline
 Cars & 0.151 & 42.96 \\ [0.5ex]
 \hline
 Trans & 0.207 & 31.11 \\ [0.5ex]
 \hline
 Utils & 0.229 & 16.79 \\ [0.5ex]
 \hline
 Rtail & 0.200 & 29.87 \\  [0.5ex]
 \hline
 Finan & 0.184 & 31.04 \\  [0.5ex]
 \hline
 Other & 0.196 & 24.94 \\ [0.5ex]
 \hline
\end{tabular}
\end{center}
\caption{Sharpe Ratio and Variance for each Industry portfolio in 17Ind}
\label{table:1}
\end{table}

Within the framework of mean-variance portfolio optimization, constructing a portfolio from these industry portfolios is expected to yield lower variance compared to the variances of individual industries. This stems from the fundamental principle of diversification, which aims to reduce risk by spreading investments across various assets. The preliminary analysis has focused on the 17Ind dataset, the smallest among the datasets available, due to the considerable size of the others making them challenging to fully detail in this context.


\newpage
\section{Out-of-Sample Evaluation: Setup}
\label{chap:setup}
\def\mathbi#1{\textbf{\em #1}}
\subsection{Methodology} 
\label{sec:gv1}
The "rolling window" procedure is used to compare the portfolio performance across time. The size of the rolling window has to be selected to perform the evaluation. Denote the rolling window length by $T$ ($T < n$), where $n$ is the total number of periods in the dataset. Following \textcite{Demiguel2009}, the estimation window of $T$ = 120 is chosen, which corresponds to 10 years of monthly data. The portfolio weights are then calculated using return data within the specified estimation window. This rolling window approach is reiterated each month, with the data for the next month being incorporated while the earliest month's data is removed, until the entire dataset has been considered. Therefore, each analysis period encompasses only 120 data points. Consequently, for every period assessed, the 132S dataset becomes singular due to its dimensions exceeding the number of observations, rendering it rank deficient. Nevertheless, we will continue to include this dataset in our assessment to demonstrate the capability of our proposed estimation approach to compute a precision matrix despite its singularity. In the end, there would be $n-T-1$ portfolio-weight vectors generated for each portfolio strategy, denoted as $w_{t}$ for $t=T, T+1,...,n-1$. The precision matrix estimators are evaluated on two factors: the condition number of the estimator and empirical portfolio metrics. The metrics used for the evaluation of the estimators will be elaborated in the preceding sections.

\subsubsection{Condition Number}
The condition number of the precision matrix measures the stability of the matrix. It is computed by taking the ratio of the largest eigenvalue to the smallest eigenvalue. To illustrate this, consider the spectral decomposition of the covariance matrix $\Sigma = U\Lambda U^{\top}$, where $\Lambda$ is a $p$ x $p$ diagonal matrix whose diagonal entries are eigenvalues of $\Sigma$, while $U$ contains the eigenvectors of $\Sigma$ corresponding to each eigenvalue, where the $i^{th}$ column of $U$ is the eigenvector corresponding to the $i^{th}$ eigenvalue in $\Lambda$. That is, $\Lambda$ = diag($\lambda_{1},..,\lambda_{p}$), where $0 < \lambda_{1} \leq ... \leq \lambda_{p}$. The spectral decomposition is as follows:

\begin{equation}
\Sigma = U\Lambda U^{\top} =\lambda_1 u_1 u_1' + \lambda_2 u_2 u_2' + \dots + \lambda_p u_p u_p'.
\end{equation}
where $u_i$ is the eigenvector corresponding to the $i^{th}$ eigenvalue. Hence the condition number for the covariance matrix is simply computed by $\frac{\lambda_{p}}{\lambda_{1}}$. The spectral decomposition of the precision matrix is given by:

\begin{equation}
\Psi = UMU' = \frac{1}{\lambda_p} u_p u_p' + \frac{1}{\lambda_{p-1}} u_{p-1} u_{p-1}' + \dots + \frac{1}{\lambda_1} u_1 u_1',
\end{equation}
where $M$ = diag($m_1, ..., m_p$) and $m_i$ = 1/$\lambda_i$. Hence the condition number for the precision matrix is computed by $\frac{m_1}{m_p}$ = $\frac{\frac{1}{\lambda_1}}{\frac{1}{\lambda_p}}$ = $\frac{\lambda_{p}}{\lambda_{1}}$. Thus the covariance matrix and the precision matrix shares the same condition number.
As can be seen from the computation, when the eigenvalues are more dispersed, the condition number will be larger. When dimension $p$ is not small relative to sample size $n$, the larger sample eigenvalues will be overestimated while the smaller sample eigenvalues will be underestimated (\textcite{Marcenko1967}, \textcite{bouchaud}). A larger condition number hence corresponds to a more unstable matrix, and the matrix is said to be ill-conditioned. For this evaluation, the condition number of the estimators will be computed for each instance of the rolling window, yielding $n-T-1$ values of condition numbers for each estimator. The mean and standard deviation of these condition numbers will then be computed. 

\subsubsection{Portfolio Metrics}
A few financial metrics are often used to evaluate the out-of-sample performance of the constructed portfolios. These metrics are illustrated below.

\paragraph{\text{Out-of-Sample Variance}}\mbox{}\\
The variance in a portfolio, which measures the dispersion of returns on assets held, acts as a barometer for gauging investment risk. When this variance is high, it signals a greater potential for fluctuation in investment returns, translating to a wider spectrum of risk. Conversely, a lower variance suggests that the investment returns are more clustered and less spread out, indicating less uncertainty in the portfolio's performance. Essentially, portfolio variance is crucial for investors as it allows them to understand and manage the unpredictability associated with their investments. A higher portfolio variance indicates greater uncertainty and risk associated with the investment.
Each rolling window instance would produce a portfolio-weight vector, denoted as $w_{t}$, and holding this portfolio for one month gives the out-of-sample return at time $t + 1$, defined as: 
\begin{equation}
R_{t} = w_{t}^{\top}r_{t+1}
\end{equation}
where $r_{t+1}$ is a vector of length $p$ of asset returns at time $t+1$. The out-of-sample mean is hence given by: 
\begin{equation}
\hat{\mu}_{w} = \frac{1}{n-T}\sum_{t=T}^{n-1}R_{t}
\end{equation}
and the out-of-sample variance can be computed as: 
\begin{equation}
\hat{\sigma}_{w}^{2} = \frac{1}{n-T-1}\sum_{t=T}^{n-1}(R_{t} - \hat{\mu_{w}})^{2}
\end{equation}

\paragraph{\text{Out-of-Sample Sharpe Ratio}}\mbox{}\\
The Sharpe ratio is a metric that evaluates an investment's or portfolio's returns in relation to its risk, essentially gauging the return per unit of risk. It is a valuable tool for investors and portfolio managers to compare the potential rewards of various investments or portfolios, factoring in the differing levels of associated risk. The out-of-sample Sharpe ratio is computed by taking the out-of-sample mean divided by the out-of-sample standard deviation, given by:
\begin{equation}
\hat{SR} = \frac{\hat{\mu}_{w}}{\hat{\sigma}_{w}}
\end{equation}
It is worth noting that the Sharpe Ratio developed by \textcite{Sharpe1963} takes into account the risk-free rate, $r_{f}$, but in this thesis we assume that $r_{f} = 0$ as we are not concerned with the performance of the portfolios in excess of the risk-free rate.

\paragraph{\text{Out-of-Sample Turnover}}\mbox{}\\
Turnover reflects the frequency at which portfolio weights undergo adjustments, serving as an indicator of the portfolio's steadiness. A portfolio with reduced turnover typically incurs fewer trading costs and suggests a greater degree of stability in the investment strategy. In practical applications, having a portfolio with many assets would result in high transaction costs, particularly if the assets are rebalanced often. These trading costs may erode the gains made by the investment strategy, indicating the importance of turnover as a metric for investors to evaluate their investment strategy. To compute the out-of-sample turnover, we would require the sequence of portfolio weights $w_{t}$. This could be computed as follows:
\begin{equation}
TO_{w} = \frac{1}{n-T-1}\sum_{t=T}^{n-1}\sum_{i=1}^{p}(|w_{i, t+1} - w_{i, t}|)
\end{equation}
where $w_{i,t}$ is the portfolio weight of asset $i$ at time $t$, and $w_{i, t+1}$ is the portfolio weight of asset $i$ at time $t+1$ after rebalancing.

\newpage
\section{Proposed Methods and Benchmarks}
For a given estimator of the precision matrix $\hat{\Psi}$, the MVP is given by: 
\begin{equation}
w_{MVP} = \frac{\hat{\Psi}e}{e^{\top}\hat{\Psi}e}
\end{equation}
Replacing $\hat{\Psi}$ with different estimators leads to different weights $w_{MVP}$ and hence different portfolio strategies. Within this section, we will examine a range of portfolio strategies and their practical applications which serve as benchmarks against which we can measure the success of our introduced techniques. 

\subsection{Benchmarks}
\subsubsection{Sample Covariance Matrix}
The MVP based on the sample covariance matrix $\mathbi{S}$ serves as the simplest alternative portfolio as it involves obtaining the inverse directly from each estimation window of the data. This portfolio will be denoted as S-MVP. 

\subsubsection{Equal-Weighted Portfolio}
The equal-weighted portfolio involves distributing the weights equally to every asset in the data, hence $w_{EW} = \frac{1}{p}e$. For instance, in the 17Ind dataset with its 17 industry portfolios, the allocation strategy would evenly divide the capital, assigning a fraction of $\frac{1}{17}$ to each portfolio. This portfolio will be denoted as EW-MVP.

\subsubsection{Ledoit-Wolf Estimator}
The estimator proposed by \textcite{LedoitWolf2004} involves computing the estimated covariance matrix as described by equation (6). As the Ledoit-Wolf estimator involves estimating a more robust estimator of the covariance matrix and not the precision matrix directly, the resulting estimator has to be inverted to obtain the precision matrix. The MVP is then obtained using the estimated precision matrix, denoted as $\hat{\Psi}_{LW}$. This portfolio will be denoted as LW-MVP.

\subsubsection{Principal Component Analysis (PCA)}
For this portfolio, the MVP is constructed based on the covariance matrix estimated by the first few principal components which can explain at least 99\% of the total variation of the data. PCA is a statistical procedure that uses an orthogonal transformation to convert a set of observations of possibly correlated variables into a set of values of linearly uncorrelated variables called principal components (\textcite{pca}). The evaluation for this portfolio is described as follows:
\begin{enumerate}
    \item For each rolling window, perform PCA on the original data to obtain the eigenvectors (principal components) and eigenvalues.
    \item Select the first $k$ principal components that explain 99\% of the variance.
    \item Compute the precision matrix, $\Psi_{PCA}$, using the selected principal components.
    \item Calculate the portfolio weights using the analytical solution $\frac{\Psi_{PCA}e}{e^{\top}\Psi_{PCA}e}$. 
    \item The portfolio weights obtained in (4) is a vector of length less than the original matrix dimension due to the dimensionality reduction. To transform the weights from the reduced dimension back to the original dimension, multiply the tranpose of the matrix of selected principal components with the weights.
    \item The weights in the original dimension are then multiplied with the returns at time $t + 1$ to obtain the portfolio returns, and subsequently the variance, Sharpe Ratio, and turnover.
\end{enumerate}
This portfolio will be denoted as PCA-MVP.

\subsubsection{No-short Sale Constraint}
The MVP based on the sample covariance matrix $\mathbi{S}$ with no-short-sale constraint proposed by \textcite{JagannathanMa2003}, denoted by JM-MVP. The no-short-sale constraint means that the weights of the portfolio must be nonnegative. This portfolio derived as follows:
\begin{equation}
\begin{aligned}
w^{\ast} = argmin_{w} \quad w^{\top}\mathbi{S} w \\
\textrm{s.t.} \quad w^{\top}e = 1\\
\quad w_{i} \geq 0, \quad i = 1,...,p
\end{aligned}
\end{equation}
Given that this portfolio strategy imposes an additional constraint $w_{i} \geq 0$, the analytical solution of the mean-variance portfolio optimization framework defined in equation (5) cannot be used to compute the weights to allocate to each asset in the data for each rolling window. 

\subsubsection{Glasso Estimator}
The MVP is based on the estimator from the glasso approach proposed by \textcite{Goto2015}. This portfolio will be denoted as Glasso-MVP. The glasso requires a regularization parameter $\rho$, which imposes the intensity of sparsity in the estimated precision matrix. For a given $\rho$, we can obtain the unique mean-variance optimizer $\hat{\Psi}_{glasso}$ even when $n > p$. Following the methodology put forth by the authors, the $\rho$ is estimated by fixing the first 120-month in-sample period before the out-of-sample testing period, and using the in-sample period to search for the value of $\rho$ that maximizes the predictive likelihood using a grid that ranges from 0 to 3, with increments of 0.1. The optimal value of $\rho$ is then kept constant through the out-of-sample period, and is not recomputed for each rolling window due to computation constraints.

\begin{table}[htbp]
\begin{center}
\begin{tabular}{||c c||} 
 \hline
 Dataset & $\rho$ \\ [1ex]
 \hline
 17Ind & 0.8 \\ [1ex]
 \hline
 30Ind & 1.0 \\ [1ex]
 \hline
 49Ind & 0.7 \\[1ex]
 \hline
 100FF & 1.1 \\[1ex]
 \hline
 132S & 0.8 \\  [1ex]
 \hline
\end{tabular}
\end{center}
\caption{Optimal $\rho$ for $\hat{\Psi}_{glasso}$}
\label{table:1}
\end{table}

Table 5 presents the best $\rho$ values identified for each dataset in the study, determined using the initial 120 data points for the $\hat{\Psi}_{glasso}$ estimator. 

\begin{figure}[htbp]
\begin{center}
\includegraphics[width=13cm]{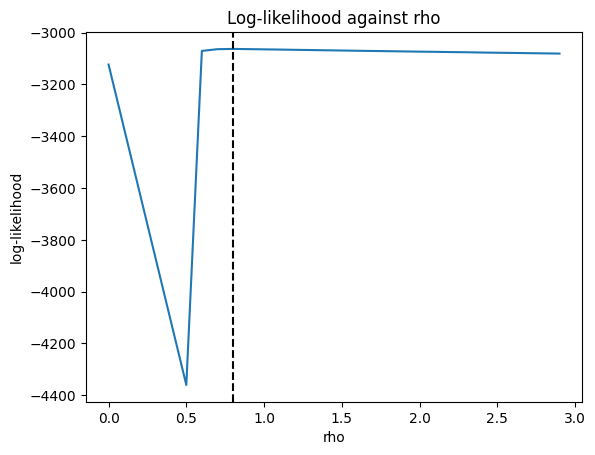}
\caption{Graph of log-likelihood value against $\rho$ for the 17Ind dataset for $\hat{\Psi}_{glasso}$}
\end{center}
\end{figure}

Figure 3 displays the optimal $\rho$ parameter for the $\hat{\Psi}_{glasso}$ estimator for the 17Ind dataset, indicated by the vertical black dotted line. This figure illustrates how the log-likelihood value changes with the penalty magnitude, and the $\rho$ which gives the highest log-likelihood value is determined to be the most optimal.

\subsection{Proposed Methods}
\subsubsection{Ridge Estimator}
Like the glasso estimator, the Ridge estimator, referred to as Ridge-MVP, also relies on a regularization parameter to control the extent of shrinkage applied to the estimated precision matrix. With a specific value of $\rho$, it is possible to derive a unique mean-variance optimizer, $\hat{\Psi}_{ridge}$, even in situations where $n > p$, by optimizing the objective function mentioned in equation (17). The method of determining $\rho$ for the Ridge estimator mirrors that for the glasso estimator, where $\rho$ is determined based on the initial 120-month in-sample period. The $\rho$ that results in the highest log-likelihood value, as outlined in equation (17), is then consistently applied across the entire dataset. Recalculating the optimal $\rho$ for every rolling window is avoided due to the high computational demand such a process would entail.

\begin{table}[htbp]
\begin{center}
\begin{tabular}{||c c||} 
 \hline
 Dataset & $\rho$ \\ [1ex]
 \hline
 17Ind & 0.4 \\ [1ex]
 \hline
 30Ind & 0.5 \\ [1ex]
 \hline
 49Ind & 0.5 \\[1ex]
 \hline
 100FF & 0.3 \\[1ex]
 \hline
 132S & 1.9 \\  [1ex]
 \hline
\end{tabular}
\end{center}
\caption{Optimal $\rho$ for $\hat{\Psi}_{ridge}$}
\label{table:1}
\end{table}

Table 6 presents the best $\rho$ values identified for each dataset in the study, determined using the initial 120 data points for the $\hat{\Psi}_{ridge}$ estimator. 

\begin{figure}[htbp]
\begin{center}
\includegraphics[width=13cm]{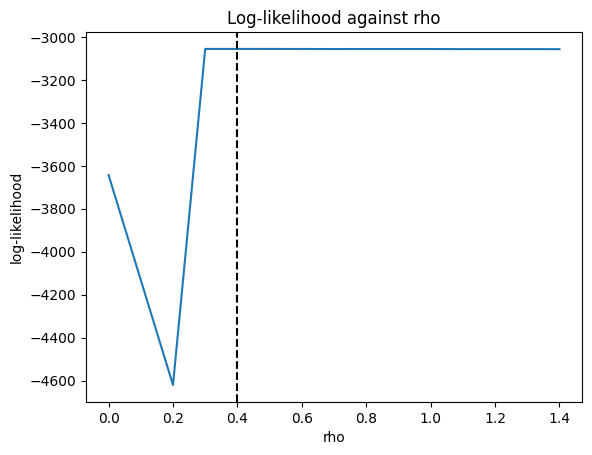}
\caption{Graph of log-likelihood value against $\rho$ for the 17Ind dataset for $\hat{\Psi}_{ridge}$}
\end{center}
\end{figure}

Similarly, Figure 4 displays the optimal $\rho$ parameter for the $\hat{\Psi}_{ridge}$ estimator for the 17Ind dataset.

\subsubsection{Elastic Net Estimator}
Similar to both the glasso and Ridge estimators, the Elastic Net estimator, hereafter referred to as EN-MVP, incorporates a regularization parameter that dictates the degree of shrinkage applied to the estimated precision matrix. This parameter combines the contributions of the $l_1$ and $l_2$ penalties in the Elastic Net approach. By setting a particular value of $\rho$, one can pinpoint a unique mean-variance optimizer, $\hat{\Psi}_{elastic}$, efficiently, even in cases where $n > p$, through the maximization of the designated objective function, as highlighted in equation (19). The process of estimating $\rho$ for the EN-MVP, akin to the glasso and Ridge estimators, involves determining it during an initial 120-month in-sample evaluation. The selected $\rho$, which yields the optimum log-likelihood as stipulated in equation (19), is consistently applied across the entire study period. Due to computational constraints, recalibrating the optimal $\rho$ for each rolling window is too inefficient.

\begin{table}[htbp]
\begin{center}
\begin{tabular}{||c c||} 
 \hline
 Dataset & $\rho$ \\ [1ex]
 \hline
 17Ind & 2.4 \\ [1ex]
 \hline
 30Ind & 1.7 \\ [1ex]
 \hline
 49Ind & 2.9 \\[1ex]
 \hline
 100FF & 0.5 \\[1ex]
 \hline
 132S & 0.6 \\  [1ex]
 \hline
\end{tabular}
\end{center}
\caption{Optimal $\rho$ for $\hat{\Psi}_{elastic}$}
\label{table:1}
\end{table}

Table 7 presents the best $\rho$ values identified for each dataset in the study, determined using the initial 120 data points for the $\hat{\Psi}_{elastic}$ estimator.

\begin{figure}[htbp]
\begin{center}
\includegraphics[width=13cm]{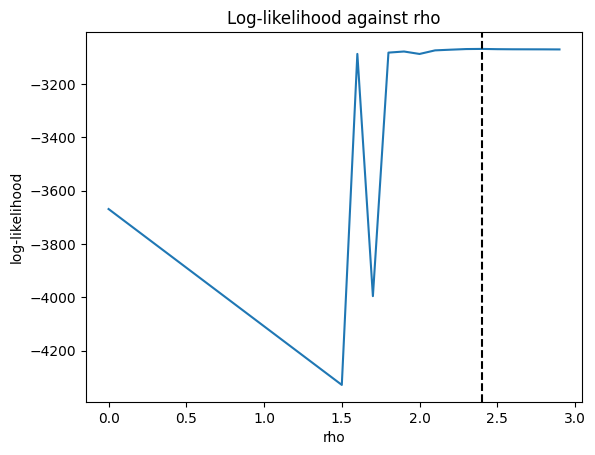}
\caption{Graph of log-likelihood value against $\rho$ for the 17Ind dataset for $\hat{\Psi}_{elastic}$}
\end{center}
\end{figure}

Likewise, Figure 5 displays the optimal $\rho$ parameter for the $\hat{\Psi}_{elastic}$ estimator for the 17Ind dataset.

In addition to the regularization parameter $\rho$, the Elastic Net estimator introduces another parameter, $\alpha$, which balances the influence of the two distinct norms. Typically, $\alpha$ is determined through a cross-validation method. Nonetheless, given the limitations in time, a simplified approach to the Elastic Net estimator has been adopted, setting $\alpha = 0.5$. This allocation implies that both the $l_1$ and $l_2$ norms are equally weighted in the objective function.

\newpage
\section{Results}
\label{chap:results}
The proposed evaluation methodology was performed on the benchmarks provided in Section 6.1, and the results were obtained for each MVP for the datasets that we have. We first report the condition numbers for different estimators of the precision matrix, which consists of $S^{-1}$, the glasso estimator $\hat{\Psi}_{glasso}$, the inverse of the Ledoit-Wolf shrunk covariance matrix $\hat{\Psi}_{LW}$, the Ridge estimator $\hat{\Psi}_{ridge}$, and the Elastic Net estimator $\hat{\Psi}_{elastic}$.

\subsection{Condition Numbers}
Table 8 reports the mean and standard deviation of condition numbers for different estimators of the precision matrix across the rolling windows. The matrix is said to be ill-conditioned if the condition number is greater than $10^{3}$ (\textcite{shi_shu_yang_he_2020}). On top of that, a matrix that is singular has a condition number equal to infinity. From Table 5, it can be seen that $S^{-1}$ is ill-conditioned for the 49Ind and 100FF datasets, and is singular for the 132S dataset, since a rolling window of size 120 means that the sample size for each window is smaller than the dimension of the dataset. On the other hand, $\hat{\Psi}_{LW}$ is not ill-conditioned for the 49Ind dataset, and its condition numbers can be computed for the 132S dataset. 

\begin{table}[htbp]
\begin{center}
  \resizebox{\textwidth}{!}{\begin{tabular}{lSSSSSSSSSSSS}
    \toprule
    \multirow{2}{*}{Dataset} &
      \multicolumn{2}{c}{17Ind} &
      \multicolumn{2}{c}{30Ind} &
      \multicolumn{2}{c}{49Ind} &
      \multicolumn{2}{c}{100FF} &
      \multicolumn{2}{c}{132S} \\
      & {Mean} & {Std. Dev} & {Mean} & {Std. Dev} & {Mean} & {Std. Dev} & {Mean} & {Std. Dev} & {Mean} & {Std. Dev}\\
      \midrule
    $S^{-1}$ & 300.60 & 145.76 & 553.58 & 220.40 & 1807.08 & 925.90  & 89007.66 & 26316.734 & {-} & {-} \\
    $\hat{\Psi}_{glasso}$ & 140.53 & 57.06 & 204.39 & 66.74 & 432.25 & 188.03 & 1144.08 & 260.23 & 953.43 & 849.11 \\
    $\hat{\Psi}_{LW}$ & 121.60 & 44.17 & 203.65 & 80.79 & 356.25 & 159.20  & 1576.79 & 549.73 & 2188.17 & 752.8 \\
    $\hat{\Psi}_{ridge}$ & 299.02 & 144.71 & 570.66 & 267.54 & 2184.43 & 1224.31  & 25904.94 & 4585.14 & 24427.94 & 4819.51 \\
    $\hat{\Psi}_{elastic}$ & 283.26 & 137.23 & 647.12 & 300.94 & 1557.01 & 661.53 & 17469.37 & 11104.55 & 48095.68 & 11778.29 \\
    \bottomrule
  \end{tabular}}
  \caption{Mean and Standard Deviation of Condition Numbers} 
\end{center}
\end{table}

The results presented in Table 8 reveal that both the $\hat{\Psi}_{glasso}$ and the $\hat{\Psi}_{LW}$ estimators surpass the performance of the sample estimator $S^{-1}$ for all evaluated datasets. This superiority is demonstrated not just by significantly lower average condition numbers across the rolling windows but also by reduced standard deviations, implying a tighter distribution of condition numbers and hence, a more consistent precision matrix. Additionally, the $\hat{\Psi}_{glasso}$ and $\hat{\Psi}_{LW}$ estimators have the capability to estimate singular matrices, as observed with the 132S dataset. In particular, the $\hat{\Psi}_{LW}$  consistently outperform the $\hat{\Psi}_{glasso}$, except for the 100FF and 132S datasets. Nonetheless, the reliability of the $\hat{\Psi}_{glasso}$ outcomes for the 100FF and 132S datasets are questionable, since a significant number of the rolling window calculations encountered errors due to the matrix being overly ill-conditioned for the base solver to handle. The estimation of $\hat{\Psi}_{glasso}$ employs the GraphicalLasso model, which is available in the open-source scikit-learn library in Python.

\begin{table}[htbp]
\begin{center}
\begin{tabular}{||c c||} 
 \hline
 Dataset & Count \\ [1ex]
 \hline
 17Ind & 389 \\ [1ex]
 \hline
 30Ind & 389 \\ [1ex]
 \hline
 49Ind & 313 \\[1ex]
 \hline
 100FF & 58 \\[1ex]
 \hline
 132S & 4 \\  [1ex]
 \hline
\end{tabular}
\end{center}
\caption{Count of estimated precision matrices from the rolling window for $\hat{\Psi}_{glasso}$}
\label{table:1}
\end{table}

Table 9 details the number of precision matrices derived through the rolling window method for each dataset, specifically for $\hat{\Psi}_{glasso}$. This procedure produced a total of 389 estimation windows, each generating a $\hat{\Psi}_{glasso}$ estimate and consequently the MVP weights. However, the counts vary by dataset: the 49Ind dataset produced 313 estimates, whereas the 100FF and 132S datasets resulted in only 58 and 4 estimates, respectively. This indicates that as the dataset's dimensionality increases, the scikit-learn GraphicalLasso model's solver struggles to compute an estimate, likely due to the matrices being excessively ill-conditioned.

Regarding the newly introduced estimators, $\hat{\Psi}_{ridge}$ and $\hat{\Psi}_{elastic}$, neither surpassed the performance of the enhanced estimators commonly referenced in the literature, such as $\hat{\Psi}_{glasso}$ and $\hat{\Psi}_{LW}$. Nevertheless, both managed to exceed the performance of the traditional sample estimator $S^{-1}$, especially in datasets with a larger number of dimensions. In the case of the 17Ind dataset, both estimators showed slight improvements over $S^{-1}$, with $\hat{\Psi}_{elastic}$ having a slight edge over $\hat{\Psi}_{ridge}$. Conversely, in the 30Ind dataset, both appeared to fall short when compared to $S^{-1}$. In the 49Ind dataset, while $\hat{\Psi}_{ridge}$ did not perform as well as $S^{-1}$, $\hat{\Psi}_{elastic}$ demonstrated superior performance. For the 100FF dataset, both estimators notably surpassed $S^{-1}$, though the resulting condition numbers still suggest that the proposed estimators may be prone to ill-conditioning. However, the introduced estimators succeeded in calculating precision matrices for singular datasets, even though their condition numbers turned out to be quite elevated. Remarkably, in the case of the largest dataset, 132S, $\hat{\Psi}_{ridge}$ notably outperformed $\hat{\Psi}_{elastic}$, achieving a mean condition number that was nearly half the size of that for $\hat{\Psi}_{elastic}$.

\subsection{Out-of-Sample Portfolio Performance}

\subsubsection{Out-of-Sample Variance}
Table 10 reports the monthly out-of-sample variances among various portfolios for various datasets. From the results, it can be seen that the various benchmarks and the proposed strategies generally outperformed the MVP with the sample covariance matrix, S-MVP, with the only exceptions being PCA-MVP for the 17Ind and 30Ind datasets, as well as the Ridge-MVP for the 30Ind and 49Ind datasets and the EN-MVP for the 30Ind dataset. EW-MVP underperformed the S-MVP consistently in terms of out-of-sample variance, up until the 100FF dataset. This is because, for the 100FF dataset, the ratio $p/n$ is very high, since $p = 100$ and $n = 120$, with the rolling horizon window being size 120. As such, the sample covariance matrix will contain large estimation errors leading to very unstable weights. The {-} on the tables indicate that the MVPs cannot be constructed because of the singularity of the covariance matrix. 

\begin{table}[htbp]
\centering
\normalsize
\begin{tabularx}{\textwidth}{lXXXXX}
  \toprule
 Strategies & 17Ind & 30Ind & 49Ind & 100FF & 132S \\ [1ex]
  \toprule
    S-MVP & 12.27 & 13.63 & 15.57 &  56.14 & {-} \\ [1ex]
    EW-MVP & 20.68 & 21.18 & 21.61 &  24.48 & 24.09 \\ [1ex]
    LW-MVP & 11.46 & 11.86 & 12.10 &  16.68 & 16.07 \\ [1ex]
    PCA-MVP & 12.37 & 13.81 & 15.02 &  18.57 & 17.26\\ [1ex]
    JM-MVP & 11.99 & 12.31 & 12.40 &  17.55 & {-} \\ [1ex]
    Glasso-MVP & 11.65 & 11.72 & 11.41 & 18.51 & 18.89 \\ [1ex]
    Ridge-MVP & 12.26 & 13.73 & 15.85 & 33.00 & 22.43 \\ [1ex]
    EN-MVP & 12.16 & 13.83 & 14.92 & 27.12 & 25.50 \\ [1ex]
   \hline
\end{tabularx}
\caption{Out-of-Sample Portfolio Risk (Monthly)} 
\end{table} 

Table 10 indicates that Ridge-MVP has a superior performance compared to S-MVP, EW-MVP, and PCA-MVP within the 17Ind dataset. In the 30Ind dataset, its superior performance extends over EW-MVP and PCA-MVP, and it surpasses only EW-MVP in the 49Ind dataset. Remarkably, within the 100FF dataset, Ridge-MVP's performance notably exceeds that of S-MVP.

In a similar vein, EN-MVP displays better outcomes than S-MVP, EW-MVP, and PCA-MVP in the 17Ind dataset. Moving to the 30Ind dataset, it only shows outperformance over EW-MVP. Within the 49Ind dataset, EN-MVP stands out against S-MVP, EW-MVP, and PCA-MVP. For the 100FF dataset, its advantage is apparent only in comparison with S-MVP.

The correlation between the condition numbers and the variances of the MVPs is noteworthy. An improvement in the condition number of an estimator over the baseline sample estimator is typically accompanied by a reduction in variance when compared to the MVP constructed using the sample estimator, known as the S-MVP. Take, for instance, the $\hat{\Psi}_{ridge}$, which presents a higher condition number than $S^{-1}$ in the 49Ind dataset, whereas the $\hat{\Psi}_{elastic}$ boasts a lower one. Correspondingly, in terms of out-of-sample portfolio variance for the 49Ind dataset, Ridge-MVP shows a higher variance relative to S-MVP, and EN-MVP demonstrates a lower variance.

\subsubsection{Out-of-sample Sharpe Ratio}
Table 11 presents the monthly Sharpe ratios obtained out-of-sample for a variety of strategies across multiple datasets. Impressively, both Ridge-MVP and EN-MVP have managed to achieve higher Sharpe ratios than the S-MVP in all datasets, with the singular exception of the 49Ind dataset where the Sharpe ratio of Ridge-MVP was marginally lower by 0.001. Even though Table 10 indicated that Ridge-MVP and EN-MVP yielded higher variances in some datasets, they still upheld superior Sharpe ratios. This can be attributed to their generation of higher returns per unit of risk taken when compared to the S-MVP. Significantly, for the 132S dataset, which is the most extensive dataset in the study, both the Ridge-MVP and EN-MVP achieved a greater Sharpe ratio than the LW-MVP. This outcome demonstrates that the estimators we introduced were able to surpass the performance of a well-established estimator commonly referenced in academic research.

\begin{table}[htbp]
\centering
\normalsize
\begin{tabularx}{\textwidth}{lXXXXX}
  \toprule
 Strategies & 17Ind & 30Ind & 49Ind & 100FF & 132S \\ [1ex]
  \toprule
    S-MVP & 0.226 & 0.246 & 0.174 &  0.159 & {-} \\ [1ex]
    EW-MVP & 0.215 & 0.217 & 0.211 &  0.212 & 0.213 \\ [1ex]
    LW-MVP & 0.266 & 0.281 & 0.218 &  0.277 & 0.274 \\ [1ex]
    PCA-MVP & 0.219 & 0.249 & 0.189 &  0.363 & 0.364 \\ [1ex]
    JM-MVP & 0.271 & 0.266 & 0.262 &  0.260 & {-} \\ [1ex]
    Glasso-MVP & 0.277 & 0.287 & 0.212 & 0.093 & 0.533 \\[1ex]
    Ridge-MVP & 0.227 & 0.246 & 0.173 & 0.227 & 0.303 \\ [1ex]
    EN-MVP & 0.232 & 0.247 & 0.177 & 0.274 & 0.280 \\ [1ex]
   \hline
\end{tabularx}
\caption{Out-of-Sample Sharpe (Monthly)} 
\end{table} 

Furthermore, excluding the 49Ind dataset, both Ridge-MVP and EN-MVP delivered superior Sharpe ratios compared to the EW-MVP across all other datasets. This highlights the advantage of using MVPs derived from estimated precision matrices over a straightforward equal-weighted portfolio approach. 

Again, it is worth noting that the Glasso-MVP gave a terrible Sharpe of 0.093 for the 100FF dataset and an inflated Sharpe of 0.533 for the 132S dataset. These results may not be reliable for the same reason as described in Section 7.1.

\subsubsection{Out-of-sample Turnover}
Table 12 reports the monthly out-of-sample turnover of different strategies for various datasets. The turnover can be interpreted as the average percentage of wealth traded across the available assets over the trading period, and reflects the stability of each portfolio. Consistent with the results of \textcite{Demiguel2009}, the EW-MVP dominates other portfolios in the sense that it provides the lowest turnover for all the datasets covered. This is unsurprising, because the EW-MVP only involves small contrarian rebalancing but does not involve any hedge trades (\textcite{Goto2015}). Not surprisingly, the turnover for S-MVP is the highest for the high-dimensional datasets. The large estimation errors in the sample covariance matrix lead to unstable portfolio weights and thus extremely high turnover (\textcite{KanandZhou2007}).
\begin{table}[htbp]
\centering
\normalsize
\begin{tabularx}{\textwidth}{lXXXXX}
  \toprule
 Strategies & 17Ind & 30Ind & 49Ind & 100FF & 132S \\ [1ex]
  \toprule
    S-MVP & 0.239 & 0.418 & 0.745 &  7.186 & {-} \\ [1ex]
    EW-MVP & 0.003 & 0.003 & 0.003 &  0.003 & 0.003 \\ [1ex]
    LW-MVP & 0.160 & 0.257 & 0.361 &  1.064 & 1.279 \\ [1ex]
    PCA-MVP & 0.247 & 0.441 & 0.751 &  1.890 & 2.037 \\ [1ex]
    JM-MVP & 0.056 & 0.067 & 0.070 &  0.116 & {-} \\ [1ex]
    Glasso-MVP & 0.153 & 0.214 & 0.329 & 0.765 & 3.114 \\ [1ex]
    Ridge-MVP & 0.238 & 0.425 & 0.760 & 3.676 & 2.832 \\ [1ex]
    EN-MVP & 0.238 & 0.421 & 0.653 & 2.884 & 3.463 \\ [1ex]
   \hline
\end{tabularx}
\caption{Out-of-Sample Turnover (Monthly)} 
\end{table} 

Table 12 indicates that for the 17Ind dataset, the Ridge-MVP and EN-MVP experienced slightly reduced turnover compared to the S-MVP. In contrast, for the 30Ind dataset, both Ridge-MVP and EN-MVP recorded higher turnover rates than the S-MVP. Within the 49Ind dataset, the turnover for Ridge-MVP was less favorable than that of the S-MVP, while the EN-MVP turnover proved to be more efficient. In the case of the 100FF dataset, both Ridge-MVP and EN-MVP displayed a notable reduction in turnover relative to the S-MVP.

Moreover, the turnover rates for both Ridge-MVP and EN-MVP were more favorable than those for PCA-MVP in the 17Ind and 30Ind datasets. Only the EN-MVP managed to achieve a lower turnover than PCA-MVP for the 49Ind dataset. However, the PCA-MVP outperformed both the Ridge-MVP and EN-MVP in terms of turnover in the 100FF dataset, achieving considerably better results.

Similar to the out-of-sample variance presented in Table 10, it is notable that enhancements in an estimator's condition number, relative to the baseline sample estimator, are generally associated with a decrease in turnover in comparison to the S-MVP. 

When assessing the performance of Ridge-MVP and EN-MVP against various benchmark strategies through different financial metrics, it becomes evident that Ridge-MVP and EN-MVP predominantly excel over S-MVP, EW-MVP, and PCA-MVP. They deliver superior results in comparison to both the conventional portfolio constructed with the sample estimator and the simple equal-weighted portfolio approach.

\newpage
\section{Discussion}

\subsection{Behaviour of Optimized Portfolio Weights}
A critical challenge within portfolio optimization is the instability of weights due to inaccuracies in estimating the covariance matrix, which can lead to extreme weight allocations. It is insightful to examine how the portfolio weights optimized by varying strategies behave in comparison to one another. Table 13 presents a summary, including the minimum, the 5th and 95th percentiles, the maximum, and the proportion of negative weights (neg) for each portfolio strategy. These metrics are averaged across all rolling windows and encapsulate the overall trend of the weight distributions. The EW-MVP strategy has been excluded because it assigns uniform weights across all assets.

In the 100FF dataset, the S-MVP exhibits substantially large weight allocations, with the portfolio weights ranging to such extremes as -0.798 for the minimum and 0.937 for the maximum. This highlights the impact that errors in estimating the covariance matrix, or more precisely, the precision matrix, can have on portfolio weights. Conversely, the LW-MVP strategy, which has consistently shown the best performance across various metrics and datasets, maintains a much narrower range of portfolio weights, with the most compact spread being from -0.170 to 0.330 for the minimum and maximum weights, respectively.

Interestingly, the range of portfolio weights for both Ridge-MVP and EN-MVP is akin to that of the S-MVP, which is unexpected given that the $l_2$ norm penalty aims to mitigate extreme weights and promote a more balanced distribution of weights. However, as shown in the table, this desired effect becomes evident primarily in datasets with larger dimensions, where the weights for both Ridge-MVP and EN-MVP appear to be more constrained and uniform in comparison to those of the S-MVP.

\begin{table}[htbp]
\centering
\begin{tabular}{l l S S S S S S S S}
\toprule
\multirow{2}{*}{Dataset} & \multirow{2}{*}{Strategies} & {Minimum} & {5\%} & {95\%} & {Maximum} & {Neg}\\
& & & & & \\
\midrule
\multirow{7}{*}{17Ind} 
& S-MVP & -0.388 & -0.259 & 0.401 & 0.606 & 0.383 \\
& LW-MVP & -0.292 & -0.182 & 0.316 & 0.489 & 0.376 \\
& PCA-MVP & -0.400 & -0.244 & 0.415 & 0.599 & 0.397 \\
& JM-MVP & 0.000 & 0.000 & 0.254 & 0.545 & 0.000 \\
& Glasso-MVP & -0.256 & -0.172 & 0.331 & 0.490 & 0.391 \\
& Ridge-MVP & -0.387 & -0.259 & 0.401 & 0.605 & 0.383 \\
& EN-MVP & -0.379 & -0.248 & 0.396 & 0.596 & 0.386 \\
\midrule
\multirow{7}{*}{30Ind} 
& S-MVP & -0.385 & -0.225 & 0.309 & 0.491 & 0.420 \\
& LW-MVP & -0.242 & -0.151 & 0.228 & 0.330 & 0.399 \\
& PCA-MVP & -0.361 & -0.209 & 0.294 & 0.507 & 0.415 \\
& JM-MVP & 0.000 & 0.000 & 0.166 & 0.459 & 0.000 \\
& Glasso-MVP & -0.206 & -0.136 & 0.214 & 0.403 & 0.402 \\
& Ridge-MVP & -0.387 & -0.226 & 0.310 & 0.493 & 0.421 \\
& EN-MVP & -0.391 & -0.230 & 0.311 & 0.510 & 0.422 \\
\midrule
\multirow{7}{*}{49Ind} 
& S-MVP & -0.367 & -0.199 & 0.249 & 0.489 & 0.458 \\
& LW-MVP & -0.170 & -0.115 & 0.172 & 0.330 & 0.429 \\
& PCA-MVP & -0.277 & -0.176 & 0.222 & 0.467 & 0.452 \\
& JM-MVP & 0.000 & 0.000 & 0.106 & 0.437 & 0.000 \\
& Glasso-MVP & -0.144 & -0.103 & 0.159 & 0.336 & 0.440 \\
& Ridge-MVP & -0.375 & -0.203 & 0.254 & 0.499 & 0.459 \\
& EN-MVP & -0.334 & -0.181 & 0.234 & 0.479 & 0.454 \\
\midrule
\multirow{7}{*}{100FF} 
& S-MVP & -0.798 & -0.501 & 0.538 & 0.937 & 0.495 \\
& LW-MVP & -0.239 & -0.156 & 0.184 & 0.308 & 0.471 \\
& PCA-MVP & -0.335 & -0.212 & 0.244 & 0.379 & 0.477 \\
& JM-MVP & 0.000 & 0.000 & 0.060 & 0.302 & 0.000 \\
& Glasso-MVP & -0.123 & -0.085 & 0.129 & 0.206 & 0.480 \\
& Ridge-MVP & -0.548 & -0.350 & 0.382 & 0.642 & 0.494 \\
& EN-MVP & -0.435 & -0.280 & 0.314 & 0.525 & 0.487 \\
\midrule
\multirow{7}{*}{132S} 
& S-MVP & {-} & {-} & {-} & {-} & {-} \\
& LW-MVP & -0.226 & -0.136 & 0.160 & 0.257 & 0.480\\
& PCA-MVP & -0.283 & -0.173 & 0.194 & 0.300 & 0.480\\
& JM-MVP & {-} & {-} & {-} & {-} & {-} \\
& Glasso-MVP & -0.094 & -0.043 & 0.077 & 0.204 & 0.417 \\
& Ridge-MVP & -0.392 & -0.238 & 0.275 & 0.459 & 0.497 \\
& EN-MVP & -0.446 & -0.268 & 0.307 & 0.526 & 0.500 \\
\bottomrule
\end{tabular}
\caption{Distribution of Portfolio Weights}
\label{tab:sharpe_ratios}
\end{table}

\subsection{Sparsity of $\hat{\Psi}_{glasso}$}
The glasso algorithm, as applied by \textcite{Goto2015}, introduces sparsity to the estimated precision matrix to address multicollinearity through feature selection. Specifically, it selects which assets to assign weights to, a process detailed in Section 3. By examining the level of sparsity in the estimated precision matrix, $\hat{\Psi}_{glasso}$, the aim is to explore the degree of multicollinearity present in the dataset. Sparsity is quantified by calculating the proportion of off-diagonal elements in $\hat{\Psi}_{glasso}$ that is zero. This calculation is performed for every estimate of $\hat{\Psi}_{glasso}$ within each rolling window and then the averages are taken across all windows to determine the overall sparsity.

\begin{table}[htbp]
\begin{center}
\begin{tabular}{||c c||} 
 \hline
 Dataset & Sparsity \\ [1ex] 
 \hline
 17Ind & 0.335 \\ [1ex] 
 \hline
 30Ind & 0.451 \\ [1ex] 
 \hline
 49Ind & 0.414 \\[1ex] 
 \hline
 100FF & 0.712 \\[1ex] 
 \hline
 132S & 0.803 \\  [1ex] 
 \hline
\end{tabular}
\end{center}
\caption{Sparsity of $\hat{\Psi}_{glasso}$}
\label{table:1}
\end{table}

Table 14 provides information on the sparsity levels of $\hat{\Psi}_{glasso}$ across various datasets. It is logical to see that sparsity increases with the dataset's dimensionality, which aligns with the understanding that multicollinearity increases as the number of variables grows. This observation is aligned with the data in Table 1, where it is noted that the maximum correlation increases alongside the dimension of the datasets.

The sparsity average varies, starting at 33.5\% for the smallest dataset and reaching up to 80.3\% for the largest dataset. This suggests that a substantial part of the precision matrix is set to zero, reflecting a strong presence of multicollinearity. Nonetheless, the reported sparsity levels for the 100FF and 132S datasets might not be fully reliable as they consist of a small portion of the original data, as elaborated on in Section 7.1 and illustrated in Table 9.

\newpage
\section{Conclusions}

In this dissertation, we embarked on an in-depth investigation into portfolio optimization, with a particular emphasis on the critical role played by the precision matrix in delineating optimal hedging strategies among stocks. It became apparent that inaccuracies in estimating this matrix could lead to the derivation of unstable portfolio weights, which, in turn, might result in suboptimal investment outcomes. In response to this challenge, we proposed a novel methodology aimed at refining the estimation of hedge portfolio weights. This approach leverages the Ridge and Elastic Net estimators to apply a penalty to the off-diagonal elements of the precision matrix.

Our comprehensive analysis has yielded compelling evidence that these refined estimators significantly mitigate out-of-sample risk, a factor of paramount importance in instances where conventional covariance matrix estimators are rendered ineffective due to their susceptibility to ill-conditioning or singularity. The empirical results obtained from our study distinctly illustrate the advantages of the proposed estimators over conventional methods in managing out-of-sample risk. Notably, portfolios optimized using our methodologies demonstrated enhanced performance metrics, such as improved Sharpe Ratios, when compared with those derived from equal-weighted and PCA-based strategies. This was particularly pronounced in the context of high-dimensional portfolios, encompassing a vast array of assets.

A salient aspect of our findings is the observed efficacy of our proposed methods in achieving significant risk reduction in portfolios characterized by high dimensionality. Herein, our methodologies not only emerged as viable alternatives but also surpassed the performance of the esteemed Ledoit-Wolf estimator, especially with respect to the largest dataset examined. This achievement corroborates the foundational premise of our research, which posits that by judiciously addressing estimation errors inherent in hedge portfolios, one can significantly amplify the efficacy of mean-variance optimizers in minimizing out-of-sample portfolio risk.

Accordingly, the contributions of this dissertation extend well beyond the realm of theoretical discourse, offering actionable insights and methodologies for portfolio managers and financial analysts grappling with the complexities of contemporary financial markets. By integrating our proposed estimation refinements, practitioners are equipped to more effectively leverage mean-variance optimization techniques, thereby ensuring more stable and reliable portfolio performance, even amidst the challenges posed by high-dimensional data.

In conclusion, this research not only enriches the academic literature on portfolio optimization but also delineates a pragmatic blueprint for enhancing portfolio management strategies in practical settings. By doing so, it signifies a notable advancement in the field of asset allocation, providing a robust foundation upon which future research and practical applications can be built.

\newpage
\section{Recommendations}
In this section, our aim is to outline potential avenues for future research and exploration that could follow from this dissertation. Given the time limitations encountered during the course of this study, we concentrated on the investigation of structure-based precision matrix estimation methods, such as the glasso, Ridge, and Elastic Net estimators, which are part of a broader family of shrinkage estimators discussed in Section 2.3.1. While structure-based estimation has received considerable attention in scholarly literature, structure-free estimation represents a more novel development. Specifically, \textcite{shi_shu_yang_he_2020} pioneered a structure-free method for estimating the precision matrix by applying constraints to the covariance matrix's eigenvalues. They demonstrated that the principal source of estimation errors lay in the sample eigenvalues rather than the eigenvectors. By applying regularization to the eigenvalues and utilizing the original eigenvectors, they were successful in developing a revised precision matrix estimator. The authors provided thorough proof of their theory and introduced an analytic solution for the calculation of the adjusted eigenvalues. Unfortunately, the exploration of structure-free approaches was beyond the scope of this thesis due to the time constraints imposed, leading to a focus on structure-based estimation methods instead.

The methodology for out-of-sample evaluation in this research incorporates a rolling window approach spanning 120 months, equivalent to a 10-year period, for precision matrix estimation. The study identifies the 132S dataset as particularly challenging due to its singularity under this framework. Empirical evidence demonstrates the superior performance of Ridge and Elastic Net estimators with datasets of larger dimensions. Two strategies are proposed for further exploration: (1) utilizing datasets with larger dimensions to test the estimators' robustness, or (2) maintaining the dataset dimensions while reducing the rolling window size, which may introduce ill-conditioning or singularity into each window's estimation process, offering a more comprehensive test for the estimators under review. 

Additionally, the practicality of using a 10-year horizon for optimal portfolio estimation is questioned, referencing studies that show market conditions undergo significant changes within such a period, necessitating adaptable investment strategies. Specifically, research utilizing the Hidden Markov Model to track market regimes over the years 2007 to 2017 reveals substantial shifts that impact investment strategy effectiveness (\textcite{regime}). Moreover, \textcite{Ang2011}'s work underscores the frequency of abrupt shifts in financial market behavior, with regime changes often tied to broad regulatory, policy, and economic shifts, thereby influencing asset prices and investor decisions.

Additionally, it is worth noting that not all assets have a decade of historical data available, as some may not have been in existence for that long. This limitation forces investors to narrow their focus to assets with over ten years of data, potentially excluding high-growth stocks that could offer substantial returns. Consequently, the second approach of shortening the rolling window period emerges as a more advantageous direction for advancing this research. This approach would not only accommodate assets with shorter historical data but also reflect the rapid changes in market conditions, thereby offering a broader, more inclusive analysis for investment decision-making.

Lastly, while it may require more computational resources, it is crucial to tailor the optimization parameter $\rho$ for the glasso, Ridge, and Elastic Net estimators individually for each rolling window instead of applying a singular value across the entire sampling period. This adjustment addresses the challenges posed by shifting market regimes. By decreasing the size of the rolling window, there is potential for more rapid computations, making it a practical approach to refine the adaptability of these estimators to dynamic market conditions. This method ensures that the estimation process remains sensitive to variations within the dataset, thereby enhancing the reliability and accuracy of the optimization outcomes. Additionally, for the Elastic Net estimator, fine-tuning the additional $\alpha$ parameter is essential to identify the ideal mix between the $l_1$ and $l_2$ norms. Due to the constraints of time, a preliminary method involving an equal mix of both norms was utilized. However, the Elastic Net estimator still demonstrated a notable advancement in performance over the baseline estimator, surpassing the improvements seen with the Ridge estimator. With meticulous optimization of $\alpha$, the Elastic Net estimator holds the promise of exceeding the performance of more existing benchmarks, indicating a significant potential for further exploration and application in optimizing portfolio performance.

\newpage
\printbibliography[heading=bibintoc]

\newpage
\appendix
\section{Appendix}
\begin{figure}[htbp]
    \centering
    \includegraphics[width=1\textwidth]{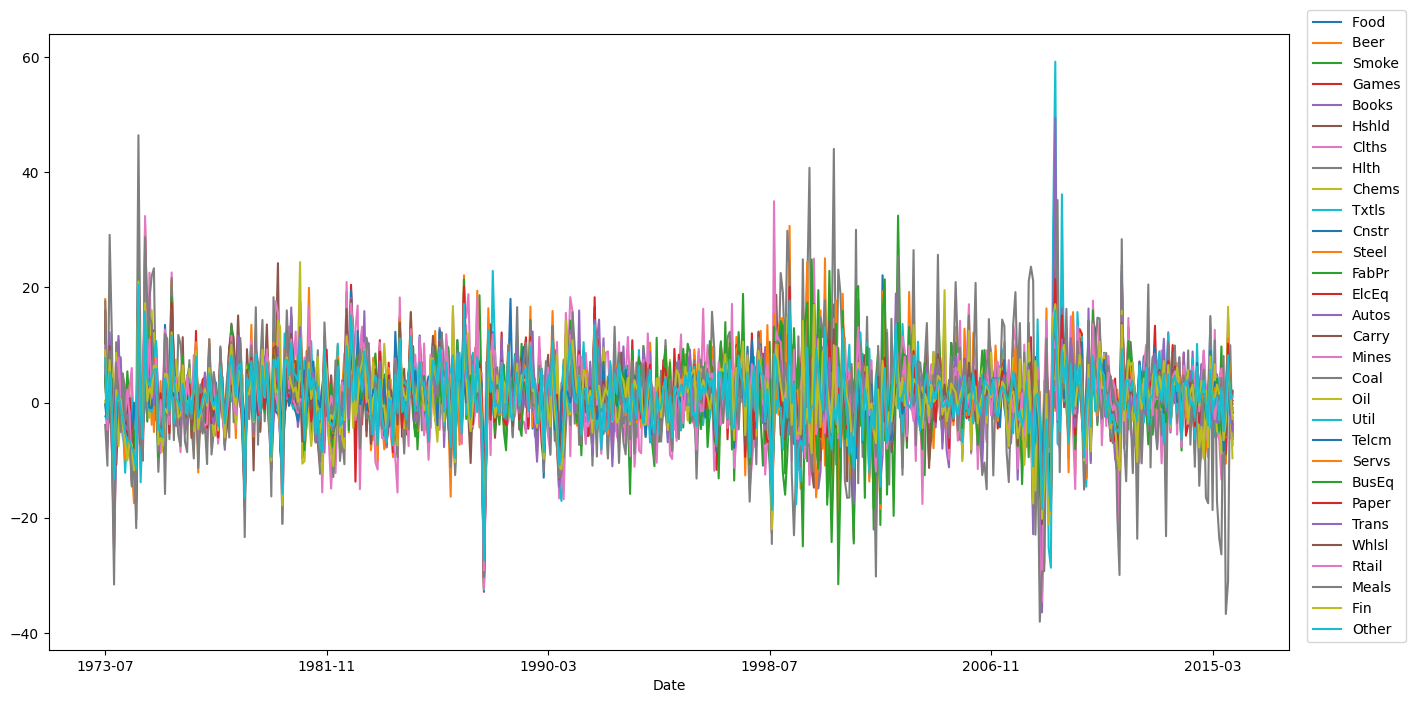}
    \caption{30Ind returns against time}
    \label{fig:gs}
\end{figure}

\begin{figure}[htbp]
    \centering
    \includegraphics[width=1\textwidth]{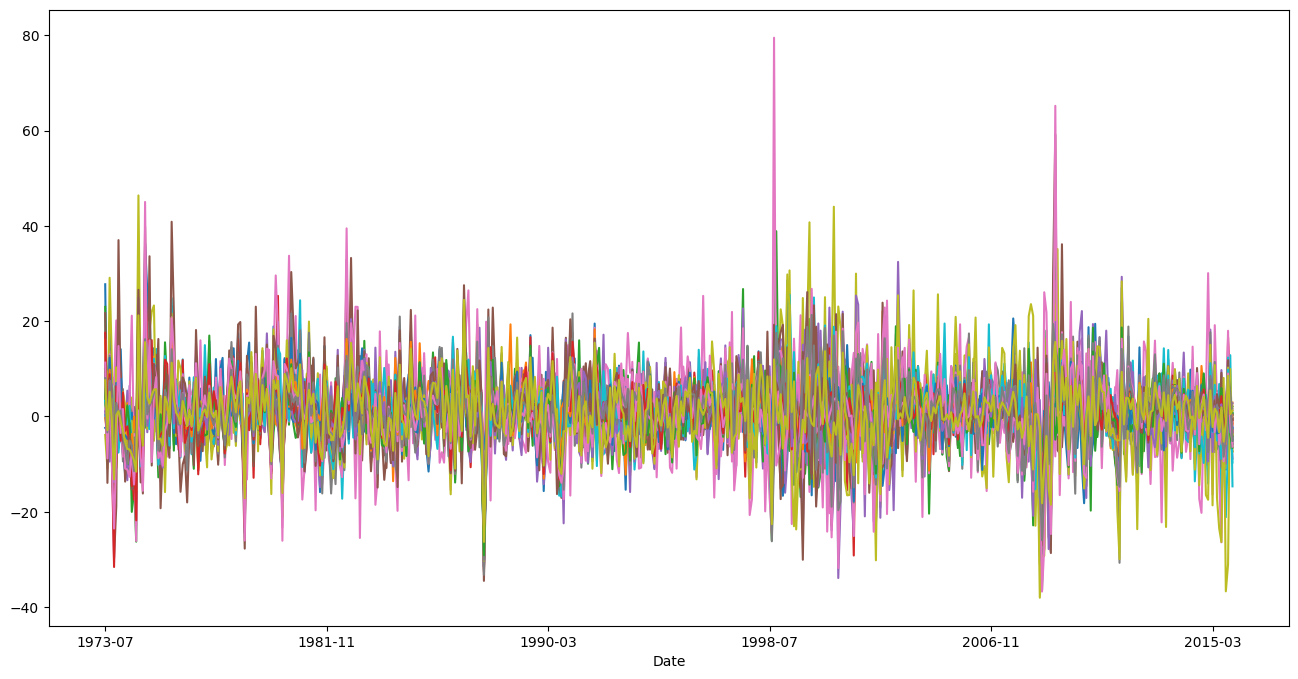}
    \caption{49Ind returns against time}
    \label{fig:gs}
\end{figure}
\begin{figure}[htbp]
    \centering
    \includegraphics[width=1\textwidth]{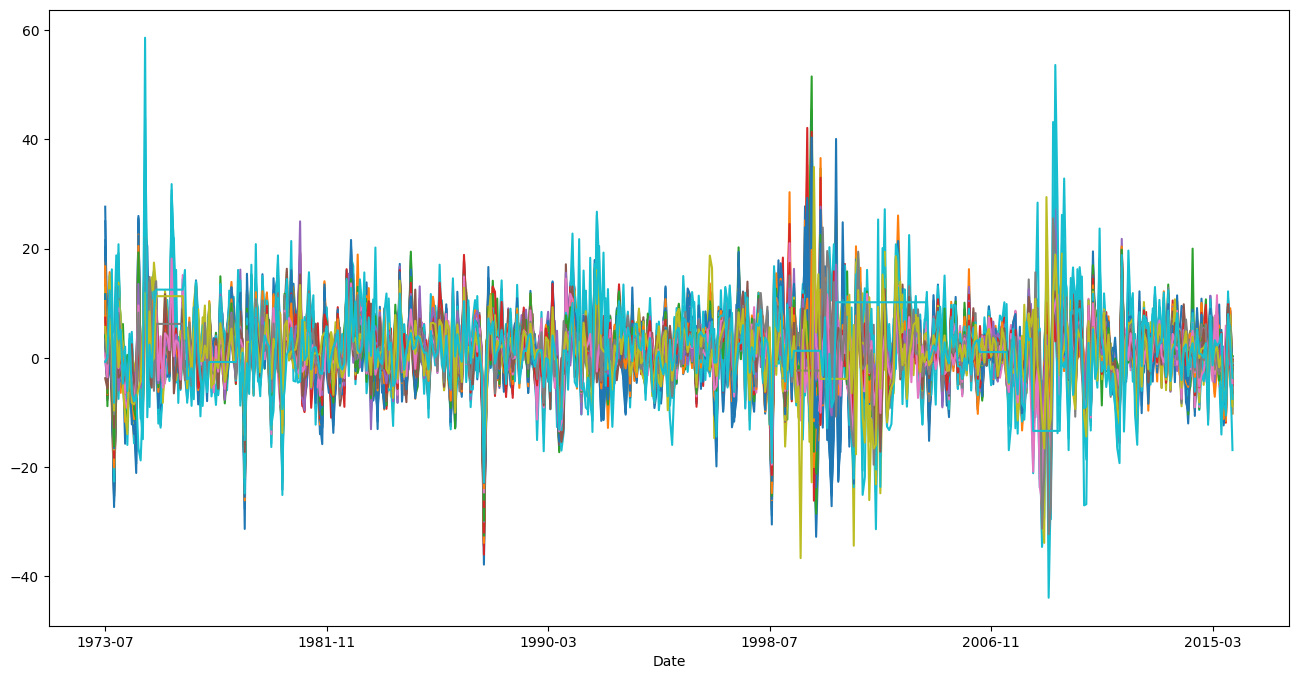}
    \caption{100FF returns against time}
    \label{fig:gs}
\end{figure}
\begin{figure}[htbp]
    \centering
    \includegraphics[width=1\textwidth]{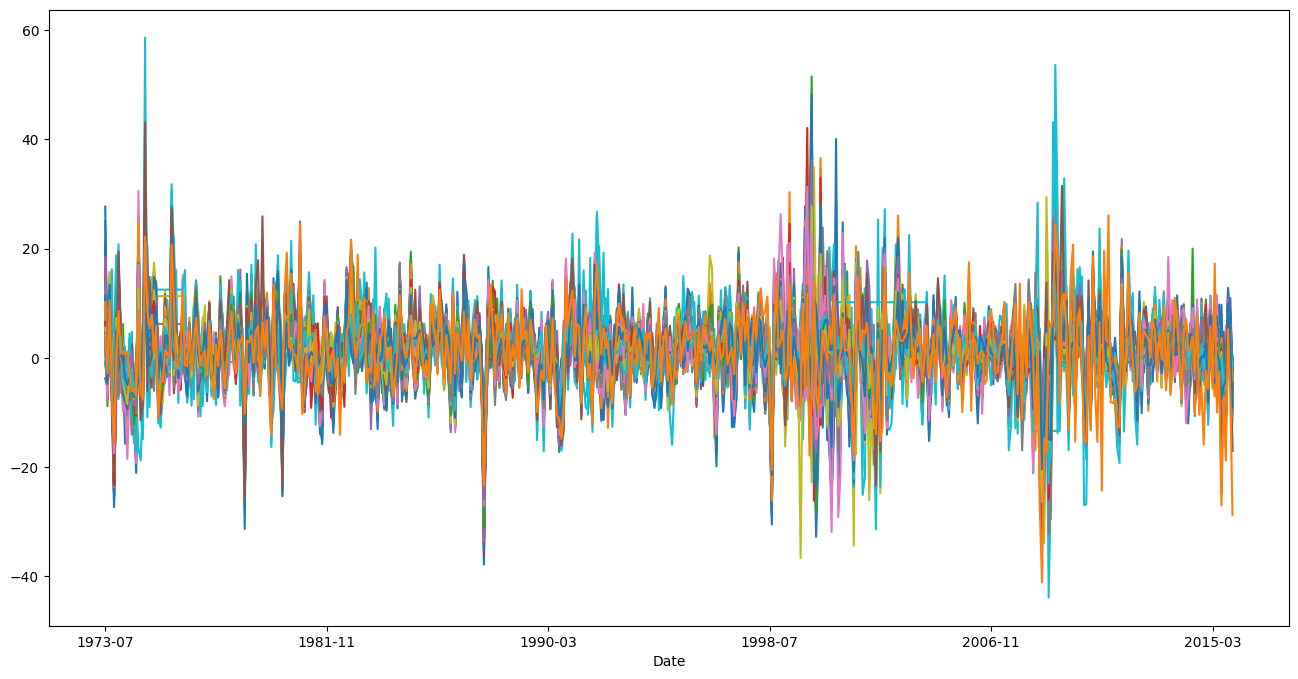}
    \caption{132S returns against time}
    \label{fig:gs}
\end{figure}

\clearpage
\newpage
\section{Appendix}
{\fontsize{16}{16}\selectfont \textbf{Derivation of Mean Variance Optimization Closed Form}}\\
By Markowitz's Mean Variance framework, the variance of a portfolio is given by:
\begin{align*}
\sigma_p^2 = \mathbf{w}^T \Sigma \mathbf{w} \quad \text{where} &\quad \Sigma = \text{covariance matrix} \\
&\quad \mathbf{w} = \text{vector of weights} \\
&\quad \mathbf{1} = \text{vector of ones} \\
&\quad \mathbf{\mu} = \text{vector of expected returns} \\
&\quad r = \text{portfolio return} \\
\\
\text{subject to} \\
&\quad \mathbf{w}^T \mathbf{1} = 1 \\
&\quad \mathbf{w}^T \mathbf{\mu} = r 
\\
\end{align*}

By introducing the Lagrangian multipliers $\lambda_1$ and $\lambda_2$ for each linear constraint, the new objective function is given by:

\begin{align*}
\quad \mathcal{L} &= \frac{1}{2} \mathbf{w}^T \Sigma \mathbf{w} - \lambda_1 (\mathbf{w}^T \mathbf{1} - 1) - \lambda_2 (\mathbf{w}^T \mathbf{\mu} - r) \\
\end{align*}

To solve for the minimum variance, differentiate $\mathcal{L}$ w.r.t to $\lambda_1$, $\lambda_2$, and $\mathbf{w}$.
\begin{align*}
\quad \frac{\partial \mathcal{L}}{\partial \mathbf{w}} &= \Sigma \mathbf{w} - \lambda_1 \mathbf{1} - \lambda_2 \mathbf{\mu} = 0 \quad &\text{(1)} \\
\frac{\partial \mathcal{L}}{\partial \lambda_1} &= \mathbf{w}^T \mathbf{1} - 1 = 0 \quad &\text{(2)} \\
\frac{\partial \mathcal{L}}{\partial \lambda_2} &= -\mathbf{w}^T \mathbf{\mu} + r = 0 \quad &\text{(3)} \\
\\
\end{align*}
From (1):
\begin{align*}
\Sigma \mathbf{w} &= \lambda_1 \mathbf{1} + \lambda_2 \mathbf{\mu} \\
\mathbf{w} &= \mathbf{\Sigma}^{-1} \mathbf{1} \lambda_1 + \mathbf{\Sigma}^{-1} \mathbf{\mu} \lambda_2 \\
\\
\end{align*}
Substituting $\mathbf{w}$ into (2) and (3):
\begin{align*}
\text{(2):} \quad & \mathbf{1}^T \mathbf{\Sigma}^{-1} \mathbf{1} \lambda_1 + \mathbf{1}^T \mathbf{\Sigma}^{-1} \mathbf{\mu} \lambda_2 = 1 \\
\text{(3):} \quad & \mathbf{\mu}^T \mathbf{\Sigma}^{-1} \mathbf{1} \lambda_1 + \mathbf{\mu}^T \mathbf{\Sigma}^{-1} \mathbf{\mu} \lambda_2 = r \\
\\
&\begin{pmatrix}
\mathbf{1}^T \mathbf{\Sigma}^{-1} \mathbf{1} & \mathbf{1}^T \mathbf{\Sigma}^{-1} \mathbf{\mu} \\
\mathbf{\mu}^T \mathbf{\Sigma}^{-1} \mathbf{1} & \mathbf{\mu}^T \mathbf{\Sigma}^{-1} \mathbf{\mu}
\end{pmatrix}
\begin{pmatrix}
\lambda_1 \\
\lambda_2
\end{pmatrix}
=
\begin{pmatrix}
1 \\
r
\end{pmatrix}
\\
\\
\text{Let} \quad 
\begin{pmatrix}
a & b \\
b & c
\end{pmatrix}
&=
\begin{pmatrix}
\mathbf{1}^T \mathbf{\Sigma}^{-1} \mathbf{1} & \mathbf{1}^T \mathbf{\Sigma}^{-1} \mathbf{\mu} \\
\mathbf{\mu}^T \mathbf{\Sigma}^{-1} \mathbf{1} & \mathbf{\mu}^T \mathbf{\Sigma}^{-1} \mathbf{\mu}
\end{pmatrix} \\
\\
\begin{pmatrix}
\lambda_1 \\
\lambda_2
\end{pmatrix}
&=
\begin{pmatrix}
a & b \\
b & c
\end{pmatrix}^{-1}
\begin{pmatrix}
1 \\
r
\end{pmatrix} \\
&=
\frac{1}{ac - b^2}
\begin{pmatrix}
c & -b \\
-b & a
\end{pmatrix}
\begin{pmatrix}
1 \\
r
\end{pmatrix}
\\
\end{align*}
Solving for $\lambda_1$ and $\lambda_2$:
\begin{align*}
\lambda_1 &= \frac{c - br}{ac - b^2} \\
\lambda_2 &= \frac{a r - b}{ac - b^2} \\
\\
\therefore \mathbf{w}^* &= \frac{-br}{ac - b^2} \mathbf{\Sigma}^{-1} \mathbf{1} + \frac{ar - b}{ac - b^2} \mathbf{\Sigma}^{-1} \mathbf{\mu}
\end{align*}
\end{document}